\documentclass[aps,preprint,showpacs,nofootinbib]{revtex4}
\usepackage{amsfonts}
\usepackage{amsmath}
\usepackage{amssymb}
\usepackage{relsize}
\usepackage{graphicx}
\usepackage{pdfpages}
\usepackage{subfigure}
\usepackage{float}
\usepackage[caption = true]{subfig}
\usepackage{color}
\usepackage{calc}
\newsavebox\CBox
\newcommand\hcancel[2][0.5pt]{%
  \ifmmode\sbox\CBox{$#2$}\else\sbox\CBox{#2}\fi%
  \makebox[0pt][l]{\usebox\CBox}%
  \rule[0.5\ht\CBox-#1/2]{\wd\CBox}{#1}}
\newcommand{\red}[1]{\textcolor{red}{#1}}

\def\fun#1#2{\lower3.6pt\vbox{\baselineskip0pt\lineskip.9pt}}
\usepackage{multirow}

\begin{document}

\title{Small recoil momenta double ionization of $\textrm{He}$ and two-electron ions by high
energy photons }

\author{M. Ya. Amusia$^{1,2}$, E. G. Drukarev$^{3}$, E. Z. Liverts $^{1}$ \\
\emph{$^1$ Racah Institute of Physics, The Hebrew University}\\
\emph{Jerusalem 91904 Jerusalem, Israel}\\
\emph{$^2$ A. F.Ioffe Physical-Technical Institute,}\\
\emph{St. Petersburg 194021,Russia }\\
\emph{$^3$National Research Center "Kurchatov Institute"}\\
\emph{B. P. Konstantinov Petersburg Nuclear Physics Institute}\\
\emph{Gatchina, St. Petersburg 188300, Russia}}


\begin{abstract}
We calculate various differential and double differential characteristics
of ionization by a single photon for  $\textrm{H}^-$, $\textrm{He}$
and for the two-electron ions with $Z=3,4,5$ in the region of the so-called quasi-free mechanism (QFM)
domination.
We employ highly accurate wave functions at the electron-electron coalescence line where coordinates of both ionized electrons coincide.
We trace the $Z$ dependence for the double differential distribution. For all considered targets we discuss the dependence of the photoelectron energy distribution on the photon energy.
Our calculation demonstrated the rapid decrease of QFM contribution with increase of the difference in energy of two outgoing electrons, and with decrease of the angle between two outgoing momenta. As a general feature, we observe the decrease of QFM contribution with nuclear charge growth.
\end{abstract}

\maketitle

\section{Introduction}

By "high energy photoionization" we mean absorption of photons with energies
$\omega $ much exceeding the single particle electron binding energies $I$,
i.e. $\omega \gg I$. If only one photoelectron is emitted,  the momentum
transferred to the nucleus that is called the recoil momentum $q$, is estimated as $q\approx p$ with $p$ being the momentum of the photoelectron. Thus, the recoil momentum strongly
exceeds the characteristic binding momentum of the ionized object $\mu
=(2mI)^{1/2}$ with $m$ being the electron mass (we employ the relativistic system of units with $\hbar $=1, $c=1$), i.e. $q\gg \mu $. This is because photoionization with only one
electron knocked out cannot take place on a free electron.

Similar situation takes place for the double photoionization, it is in
emission of two electrons by a single photon, while the photon energy $%
\omega $ is not too large. The sharing of energy is strongly unequal and $%
q\approx p_{1}\approx (2m\omega )^{1/2}$ with $p_{1}$ standing for momentum
of the faster photoelectron, while the second electron is emitted with
momentum $p_{2}\sim \mu $. However, with the increase of $\omega $
the role of so-called quasi-free mechanism (QFM) suggested in \cite{1}
becomes more and more important. In the frame of QFM momenta of
photoelectrons $\mathbf{p}_{1,2}$ and that of the photon $\mathbf{k}$
compose such configuration that the recoil momentum
\begin{equation}
\mathbf{q}=\mathbf{k}-\mathbf{p}_{1}-\mathbf{p}_{2},  \label{1}
\end{equation}%
becomes as small as the binding momentum $\mu $, i.e.
\begin{equation}
q\sim \mu .  \label{2}
\end{equation}%
Since each act of transfer of large momentum $q\gg \mu $ to the nucleus
leads to the small factor $1/q^{2}$ in the amplitude, the QFM provides
surplus in differential characteristics and in the total cross section of double
ionization.

Most of the publications on QFM touched the theory of the mechanism for the
case of two-electron (helium-like) atomic systems. In the first calculation of the QFM
contribution to the total double photoionization cross section \cite{2} it
was shown to become the main mechanism of the process at the energies of
hundreds keV. The nuclear charge dependence of QFM contribution to the cross
section was traced in \cite{3}. It was emphasized in \cite{4} that the
description of the QFM requires employing the two-electron bound state wave
function $\psi (\mathbf{r}_{1},\mathbf{r}_{2})$ with the proper analytical
behavior on the electron-electron coalescence line $\mathbf{r}_{1}=\mathbf{r}_{2}$.
It should satisfy
the known relation between the precise two-electron wave function and its coordinate derivative at the line of zero interelectron distance known as
the second Kato condition \cite{5}.
The fulfillment of  this condition by usually employed in calculations \red{of} approximate wave functions is necessary  for proper description of QFM since it accounts for the singularity of the Coulomb interlectron interaction.
The change of the
spectrum curve caused by the QFM with the growth of the photon energy was
analyzed in \cite{6}. For more details and references see Ch.9 of the book
\cite{7}.

It follows from the works mentioned above that for helium the QFM provides a
noticeable contribution to the spectrum of photoelectrons starting from the
photon energies of about 2\thinspace keV. The QFM corrections to the total
cross section become noticeable at the photon energies of dozens \thinspace keV.
For heavier two-electron ions the corresponding photon energies become
larger. As it stands now, experimental data for such energies are not
available.

Although the QFM was discussed in literature during many years, to detect it
experimentally remained a challenge till this was done by the group of D\"{o}%
rner \cite{8}. Note that this discovery was made 38 years after its
prediction \cite{1} and became possible only after invention of a new
experimental technique which enables investigation of the double electron photoionization
as a function of recoil momentum $q$. However, the obtained clear
manifestation of the QFM has been detected at much smaller value of the
photon energy $\omega =800$\thinspace eV than expected. It was found in \cite%
{8} that the distribution in momenta $q$ transferred to the final state
doubly charged ions in double photoionization of helium has a surplus at
small $q$ of the order of $1-2$ atomic units. However \cite{8} did not
contain quantitative results.

An important move mainly in experimental investigation and not only of
helium atom but of hydrogen molecule also has been made by the quite recent
publication \cite{8a}. The development of experimental technique leaves no
doubt that investigation of the double electron ionization by a single
photon as a function of recoil momentum becomes an important tool in studies
of short-range interelectron correlations in atoms, molecules and, perhaps,
more complex compounds.

In \cite{8a} the double differential distributions {$d^2\sigma /d\tau d\beta $}
with $\tau =\mathbf{p}_{1}\mathbf{p}_{2}/p_{1}p_{2}$ and
\begin{equation}
\beta =\frac{|\epsilon _{1}-\epsilon _{2}|}{E},  \label{3}
\end{equation}%
($\varepsilon _{1,2}$ are the
energies of the photoelectrons, $E=\varepsilon _{1}+\varepsilon _{2}=\omega -I^{++}$, where $I^{++}$ is the two-electron ionization potential)
have been measured for $\textrm{He}$ atom and $\textrm{H}_2$ molecule. Quite a powerful QFM peak at
$\tau =-1$, $\beta =0$ have been observed in both objects.  Also the peak in the
energy distribution
\begin{equation*}
\frac{d\sigma}{d\beta} =\int_{(p_1-p_2)^2}^{(p_1+p_2)^2}\left(\frac{d^2\sigma}{dq^2d\beta}\right)d(q^2)
\end{equation*}
for the same targets attributed to
QFM have been seen.

These results prompt theoretical investigation of QFM for other, not yet
investigated two-electron systems that can become the objects of
photoionization studies soon. Note that the results of \cite{8} stimulated us to calculate
the differential distributions of the process for $\textrm{He}$\ atom at $q\sim
\mu $ and photon energies $\omega \approx 1$\thinspace keV \cite{9}. In
\cite{9} we employed approximate bound state wave functions at the
electron-electron coalescence line obtained in the work \cite{10}. This enabled us to carry
out analytical calculations.

Since then the ability to calculate improved considerably.
So, in the present paper, we employ much more sophisticated bound
state wave functions \cite{11},\cite{12} having in mind the impressive increase in experimental
accuracy achieved in \cite{8a}. We also extend our calculations to
include all the lightest two-electron positive ions ($Z\leq 5$) and the negative
hydrogen ion $\textrm{H}^{-}$. For $\textrm{He}$\ atom we trace the dependence of the
double differential distributions on the photon energy $\omega $.
Including several two-electron ions, we
trace the $Z$ dependence of the double differential contributions for photon
energies around 1 keV.

While we consider the photon energies corresponding to nonrelativistic
photoelectrons, i.e. $\omega \ll m$, the QFM is possible only in the
vicinity of the center of the energy distribution, where the relative
difference of the electron energies $\beta$
is small, $\beta \ll 1$. The actual value of $\varepsilon _{1}-\varepsilon
_{2}$ where the QFM is possible depends on the ratio $k/\mu $ of the photon
momentum $k=\omega $ and of the characteristic moment of the bound state $%
\mu $ \cite{1}. We consider the case $\omega \ll \mu $. For helium this
means $\omega \ll 6$ keV. In this case $p_{1}-p_{2}\leq q$ and the QFM
is at work if
\begin{equation}
\beta \leq \sqrt{\frac{q^{2}}{mE}} \label{4}
\end{equation}%
Condition (2) requires also that the photoelectrons move in approximately
opposite directions since $\tau =\mathbf{p_{1}}\mathbf{p}%
_{2}/p_{1}p_{2}=(q^{2}-p_{1}^{2}-p_{2}^{2})/(2p_{1}p_{2})\approx -1$.

An important feature of the QFM is that its amplitude $F$ can be expressed
in terms of the amplitude $F_{0}$ that represent moving to continuum due to
the photon absorption by two free electrons at rest-see below and \cite{7}.
This explains the name "quasifree"- the two-electron system cam move almost
without noticing the nucleus. However to do this the motion of the electrons
should be strongly correlated.

One can see that the QFM is impossible in the dipole approximation where we
must put $\mathbf{k}=0$. Thus the photoelectrons move exactly bach-to-back
with $\mathbf{p}_{1}+\mathbf{p}_{2}=0$. The incoming photon carries spin $%
S=1 $ while the two-electron system in spin singlet state can not carry
angular momentum $J=1$. Thus, we must include the quadrupole terms of
interaction between the photon and electrons.

Presenting $\varepsilon _{2}=(\mathbf{p}_{1}-\mathbf{q})^{2}/2m$ we find for
the differential cross section corresponding to the QFM
\begin{equation}
d\sigma \ =\ 2\pi \delta \Big(E-2\varepsilon _{1}-\frac{p_{1}q_{z}}{m}-\frac{%
q^{2}}{2m}\Big)|F|^{2}\frac{d^{3}p_{1}}{(2\pi )^{3}}\frac{dq^{2}dq_{z}}{4\pi
}.  \label{5}
\end{equation}%
Here $F$ is the amplitude describing the QFM mechanism; the averaging over
photon polarizations should be carried out. Also, $z$ is the direction of
momentum $\mathbf{p}_{1}-\mathbf{k}$, and we put $\mathbf{p}_{1}-\mathbf{k}\simeq%
\mathbf{p_{1}}$ in the argument of the delta-function. Using the
delta-function for integration over $q_{z}$ we obtain for the energy
distribution
\begin{equation}
\frac{d^2\sigma }{dq^{2}d\beta}\ =\ \frac{m^{2}E}{2}\int \frac{\left\vert
F\right\vert ^{2}dt}{(2\pi )^{3}};\quad t\ =\ \mathbf{p}_{1}\mathbf{k}%
/p_{1}k,  \label{6}
\end{equation}%
Another double differential distribution of interest is
\begin{equation}
\frac{d^2\sigma }{dq^{2}d\tau }\ =\ 2p_{1}p_{2}\frac{d^2\sigma }{dq^{2}d\beta }%
=m^{2}E^{3}\int \frac{\left\vert F\right\vert ^{2}}{(2\pi )^{3}}{dt}.
\label{7}
\end{equation}%
Employing these expressions, one can obtain other differential
distributions, e.g.
\begin{equation}
\frac{d\sigma }{dq^{2}}\ =\int_{0}^{q/p}d\beta \frac{d^2\sigma }{dq^{2}d\beta }%
;\quad p=(mE)^{1/2}.  \label{8}
\end{equation}

\section{The QFM amplitude}

We introduce
\begin{equation}
\mathbf{R} = (\mathbf{r}_1+\mathbf{r}_2)/2; \quad {{\mbox{\boldmath$\rho$}}}
=\mathbf{r}_1-\mathbf{r}_2,  \label{9}
\end{equation}
with $\mathbf{r}_{1,2}$ denoting the positions of the two electrons in the
rest frame of the nucleus. We present the ground state wave function in
terms of these variables
\begin{equation}
\Psi(\mathbf{r}_1, \mathbf{r}_2)\ =\ \hat\Psi(\mathbf{R}, {{%
\mbox{\boldmath$\rho$}}})\,.  \label{10}
\end{equation}

It is instructive to start with the QFM amplitude $F^{(0)}$ in which the
photoelectrons are described by the plane waves. Thus, the wave function of
the photoelectrons is
\begin{equation}
\Psi _{ph}(\mathbf{r}_{1},\mathbf{r}_{2})\ =\frac{1}{\sqrt{2}}\Big(\psi
_{p_{1}}(\mathbf{r_{1}})\psi _{p_{2}}(\mathbf{r_{2}})+\psi _{p_{1}}(\mathbf{%
r_{2}})\psi _{p_{2}}(\mathbf{r_{1}})\Big),  \label{10a}
\end{equation}%
with $\psi _{p_{j}}(\mathbf{r})=e^{-i\mathbf{p}_{j}\mathbf{r}}$.
Analysis that employs such a wave function contains all essential physics.

Introducing ${{\mbox{\boldmath$\kappa$}}}=(\mathbf{p}_{1}-\mathbf{p}%
_{2})/2\approx \mathbf{p}_{1}$ we write
\begin{equation}
F^{(0)}=\sqrt{2}N(\omega )\int d^{3}Rd^{3}\rho e^{-i\mathbf{q}\mathbf{R}+i({{%
\mbox{\boldmath$\kappa$}}}-\mathbf{k}/2) \mathbf{r}}\Big(\frac{i\mathbf{%
e}\cdot {\nabla }_{\rho }}{m}-\frac{i\mathbf{e}\cdot {\nabla }_{R}}{2m}\Big)%
\hat{\Psi}(\mathbf{R},\mathbf{r})+(\mathbf{p}_{1}\leftrightarrow \mathbf{p}%
_{2}),  \label{11}
\end{equation}
Here ${\bf e}$ is the photon polarization vector, $N(\omega )=\sqrt{4\pi \alpha /2\omega }$ is the normalization factor of
the photon wave function, while $\alpha\simeq1/137$ is the fine structure constant. Integrating by parts we find that since $\kappa =|{%
{\mbox{\boldmath$\kappa$}}}|\gg q$, the first term in the parenthesis on the
right hand side dominates, providing
\begin{equation}
F^{(0)}=\sqrt{2}N(\omega )\frac{{\mathbf{e}{{\mbox{\boldmath$\kappa$}}}}}{m}%
\int d^{3}Rd^{3}\rho e^{i\mathbf{q}\mathbf{R}+i({{\mbox{\boldmath$\kappa$}}}-%
\mathbf{k}/2) \mathbf{r}}\hat{\Psi}(\mathbf{R},\mathbf{r})+(\mathbf{p}%
_{1}\leftrightarrow \mathbf{p}_{2}).  \label{12}
\end{equation}%
The integral is determined by $R\sim 1/q\sim 1/\mu $  i.e. the
characteristic $R$ are of the order of the size of the bound state. The
important values of $\rho $ are much smaller being of the order $1/\kappa
\ll 1/\mu$. To pick the quadrupole terms we present the wave function
as
\begin{equation}
\hat{\Psi}(\mathbf{R},{{\mbox{\boldmath$\rho$}}})\ =\ \hat{\Psi}%
(R,0,0)+\zeta \hat{\Psi}^{\prime }(R,\zeta ,0)|_{\zeta =0}+\rho \hat{\Psi}%
^{\prime }(R,0,\rho )|_{\rho =0}+0(\rho ^{2}),  \label{13}
\end{equation}%
with $\zeta ={\mathbf{R} {{\mbox{\boldmath$\rho$}}}}$. Substituting this
expansion into the integral over $\mathbf{\rho }$ in Eq.(\ref{11})
\begin{equation}
J(\mathbf{a},R)=\int d^{3}\rho e^{i\mathbf{a} {{\mbox{\boldmath$\rho$}}}%
}\hat{\Psi}(\mathbf{R},\mathbf{r}),  \label{14}
\end{equation}%
with
\begin{equation}
\mathbf{a}=\frac{\mathbf{p}_{1}-\mathbf{p}_{2}-\mathbf{k}}{2}  \label{14a}
\end{equation}%
we see that only the third term on the right hand side of Eq.(\ref{13})
contributes, providing
\begin{equation}
J(\mathbf{a},R)=-\frac{8\pi \hat{\Psi}^{\prime }(R,0,\rho )|_{\rho =0}}{a^{4}%
}=-\frac{4\pi m\alpha }{a^{4}}\hat{\Psi}(\mathbf{R},0).  \label{15}
\end{equation}%
The second equality is due to the second Kato cusp condition \cite{5}
\begin{equation*}
\frac{\partial \hat{\Psi}(\mathbf{R},{{\mbox{\boldmath$\rho$}}})}{\partial \rho}|_{\rho
=0}=m\alpha \hat{\Psi}(\mathbf{R},{\mbox{\boldmath$\rho$}}=0)/2.
\end{equation*}%
Thus, the amplitude
\begin{equation}
F^{(0)}\ =\ \sqrt{2}N(\omega )\frac{{\mathbf{e}{{\mbox{\boldmath$\kappa$}}}}}{m%
}\int d^{3}R~e^{i\mathbf{q}\mathbf{R}}J(\mathbf{a},R)+\Big(\mathbf{p}%
_{1}\leftrightarrow \mathbf{p}_{2}\Big)  \label{16}
\end{equation}%
can be written as
\begin{equation}
F^{(0)}=F_{0}S(q).  \label{17}
\end{equation}%
Here
\begin{equation}
S(q)\ =\int d^{3}re^{i\mathbf{q}\mathbf{r}}\Psi (\mathbf{r},\mathbf{r})\
=\int \frac{d^{3}f}{(2\pi )^{3}}\tilde{\Psi}(\mathbf{q}-\mathbf{f},\mathbf{f}%
)  \label{18}
\end{equation}%
describes transfer of momentum $\mathbf{q}$ from the nucleus to the bound
electrons. In the lowest order of expansion in powers of $I^{++}/\omega $ we
put $E=\omega $, and as a result have
\begin{equation}
F_{0}\ =\ -4\pi \sqrt{2}\alpha N(\omega )\frac{{\mathbf{e}{{
\mbox{\boldmath$\kappa$}}}}}{a^{4}}+\Big(\mathbf{p}_{1}\leftrightarrow
\mathbf{p}_{2}\Big),  \label{19}
\end{equation}%
the amplitude of the process in which one photon moves the system
consisting of two free electrons in spin-singlet state to continuum.

In the lowest (dipole) approximation we must put $\mathbf{k}=0$ in the
factor $1/a^{4}$ with $a$ defined by Eq.(\ref{14a}). This leads to $F_{0}=0$
and $F^{(0)}=0$ in agreement with the analysis presented above. The leading
nonvanishing contribution is provided by next to leading term of expansion
of the factor
\begin{equation}
\frac{1}{a^{4}}=\frac{1}{m^{2}E^{2}}\left( 1+\frac{2\mathbf{p_{1}}\mathbf{k}%
}{mE}\right) .  \label{19a}
\end{equation}%
Thus the amplitude of the process on the free electrons is
\begin{equation}
F_{0}\ =\ -16\pi \sqrt{2}\alpha N(\omega )\frac{(\mathbf{e}\mathbf{p}_1)(%
\mathbf{p}_{1}\mathbf{k})}{m^{3}E^{3}},  \label{20}
\end{equation}%
while the amplitude for the process on the bound electrons is given by Eq.(%
\ref{17}).

Now we describe the photoelectrons by nonrelativistic Coulomb functions. Note
that we do not employ expansion in powers of $I^{++}/\omega $. The two-electron wave function is presented by Eq.(\ref{10a}) with
\begin{equation}
\psi _{p_{j}}(\mathbf{r})=e^{-i\mathbf{p}_{j}\mathbf{r}}X_{p_{j}}(\mathbf{r}%
);\quad X_{p_{j}}(\mathbf{r})=N(p_{j})_{1}F_{1}(i\xi _{j},1,ip_{j}r-i
\mathbf{p}_j\mathbf{r}),  \label{21}
\end{equation}%
where $_{1}F_{1}(b,1,z)$ is the confluent hypergeometric function of the first kind, $\xi _{j}=m\alpha Z/p_{j}$, $%
N(p_{j})=2\pi \xi _{j}/(1-e^{-2\pi \xi _{j}})=\psi _{p_{j}}(\mathbf{r}=0)$.
Evaluation similar to that carried out for the case when the photoelectrons
are described by plane waves \cite{9}, \cite{7} provides
\begin{equation}
F\ =\ F_{0}S_{1}(q)\,.  \label{22}
\end{equation}%
Here $F_{0}$ is given by Eq.(\ref{20}) while
\begin{equation}
S_{1}(q)\ =\int d^{3}Re^{i\mathbf{q}\mathbf{R}}X_{p_{1}}(\mathbf{R}%
)X_{p_{2}}(\mathbf{R})\tilde{\Psi}(R,0),  \label{23}
\end{equation}%
see Eqs. (\ref{17}), (\ref{18}).
The corrected analytical representation for the integral $S_1(q)$ is presented in the Appendix.

\section{Differential distributions}

Combining Eqs.(\ref{6}), (\ref{22}) and (\ref{23}) we find for the double
differential distribution
\begin{equation}
\frac{{d^2}\sigma}{dq^{2}d\beta }=\frac{2^{6}}{15}\alpha ^{3}\frac{\omega
|S_{1}(q)|^{2}}{m^{2}E^{3}}.  \label{24}
\end{equation}%
Note that the photon energy $\omega$ is much larger than the two-electron ionization energy $I^{++}$.
Therefore  we used the approximation $E\simeq \omega$ in the real calculations.

In Figs.1 and 2 we trace the $Z$ dependence of the distributions $d^2\sigma/dq^2d\beta$ and
$d\sigma/dq^2$ correspondingly.
We present the results for $\textrm{He}$ as well as for $\textrm{H}^-$ and the two-electron ions of the nuclei with $Z=3,4,5$. These distributions were studied in \cite{8} for $\textrm{He}$ only.
The horizontal axis is for $q^2$. The vertical axis is for $d\sigma/dq^2d\beta$ in $barns \cdot a_0^2$ and for
$d\sigma/dq^2$ in barns.
In Figs. 1a and 2a we show these distributions for helium at $\omega=800$ eV (the studies in \cite{8} were  carried out for this value of $\omega$).
To make comparison for different ions more sensible, for other objects the energies were changed proportionally to the total binding energy $I^{++}$,
that is 0.53, 2.90, 7.28, 13.66, and 22.03 for Z=1, 2, 3, 4, and 5, respectively.
To have a feeling of dependence of these distributions on the photon energy $\omega$ we present them for $\omega=1000$ eV (in Figs.1b and 2b, in \cite{8}) were  carried out for this value of $\omega$)
in $\textrm{He}$, and energies for other objects modified accordingly to their respective values of $I^{++}$.
In these figures as well as in Figures 3,4 we change the values of $\omega$ for $\textrm{H}^-$, ($Z=1$), $\textrm{Li}^{+}$ ($Z=3$), $\textrm{Be}^{2+}$ ($Z=3$) and $\textrm{B}^{3+}$ ($Z=5$)  proportionally to the total binding energy $I^{++}$,
as compared to that of $\omega=800 (1000)$ eV for $\textrm{He}$.

\begin{figure}[htbp]
\centering
\caption{Distribution $d^2\sigma/d(q^2)d\beta$ in $10^{-10} a_0^4$ is presented as a function of $q^2$ in $a_0^{-2}$, where
$a_0$ is the Bohr radius, and $\beta=0$. The solid lines correspond to the photon energies $\omega=145,800,2000,3750,6100$ eV, whereas the dashed lines correspond to the photon energies $\omega=180,1000,2500,4700,7600$ eV for $\textrm{H}^-$, $\textrm{He}$ and helium-like ions with $Z=3;4;5$, respectively.}
\subfigure{\includegraphics[width=0.495\textwidth]{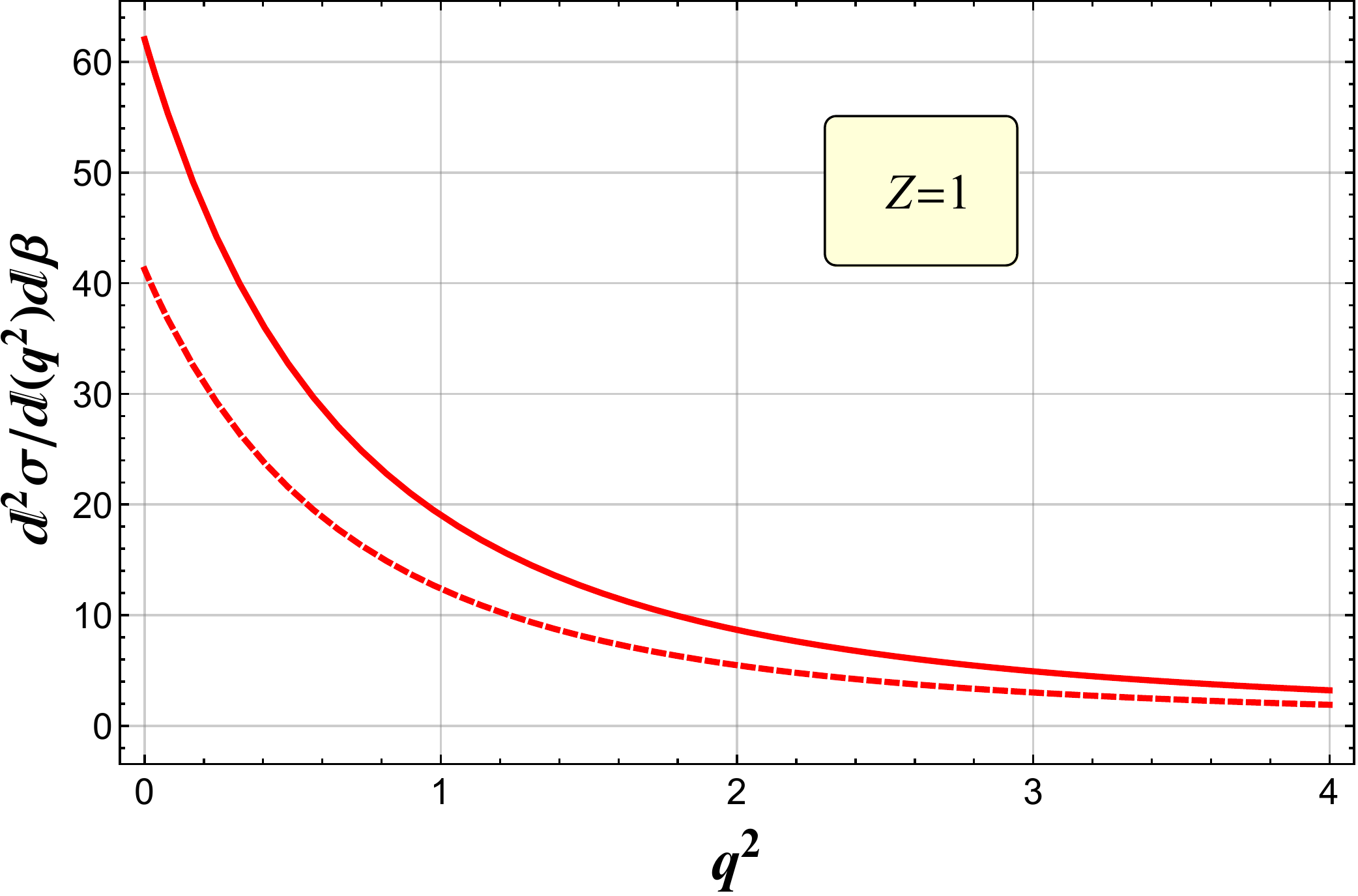}}\label{F1:a}
\subfigure{\includegraphics[width=0.495\textwidth]{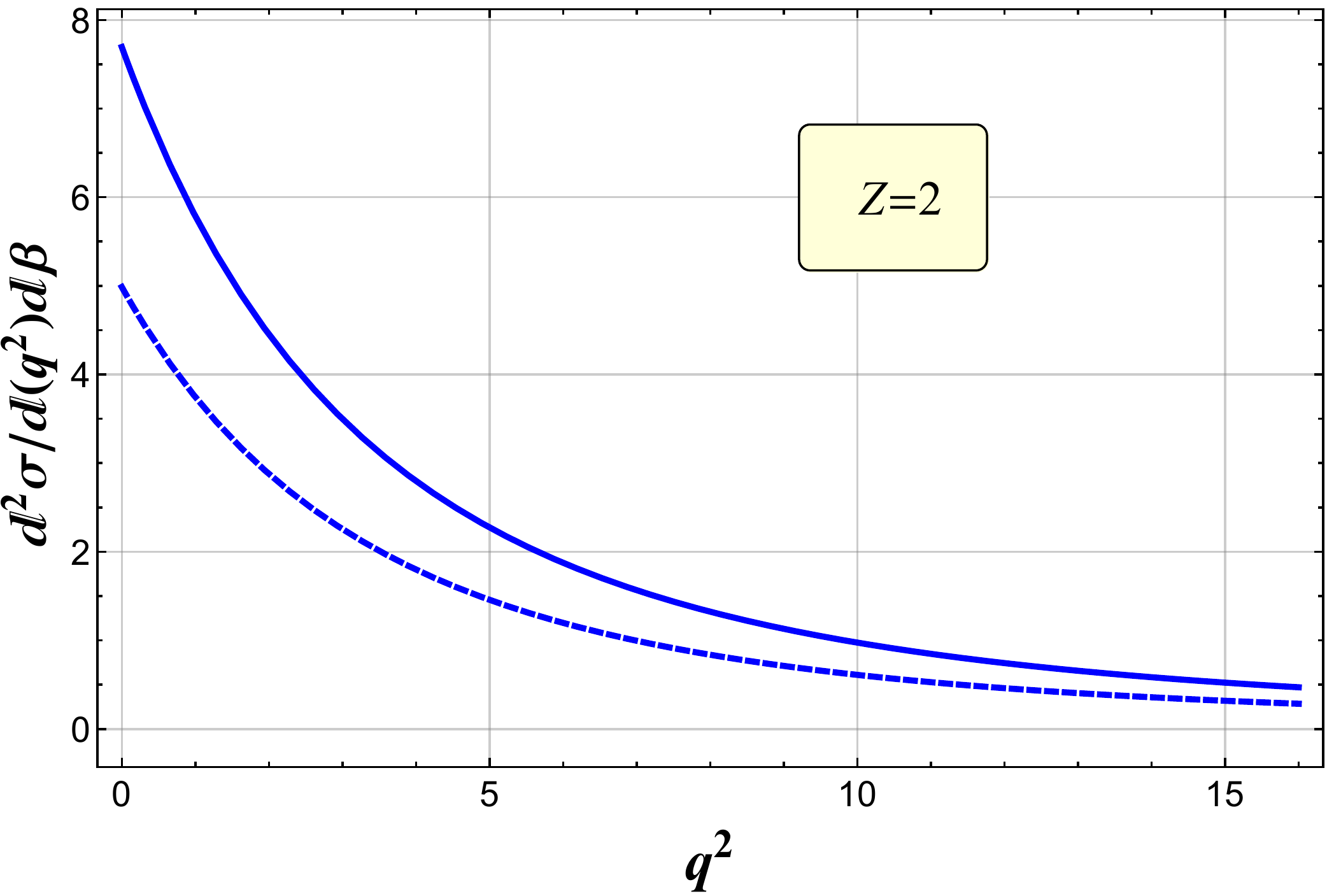}}\label{F1:b}
\subfigure{\includegraphics[width=0.495\textwidth]{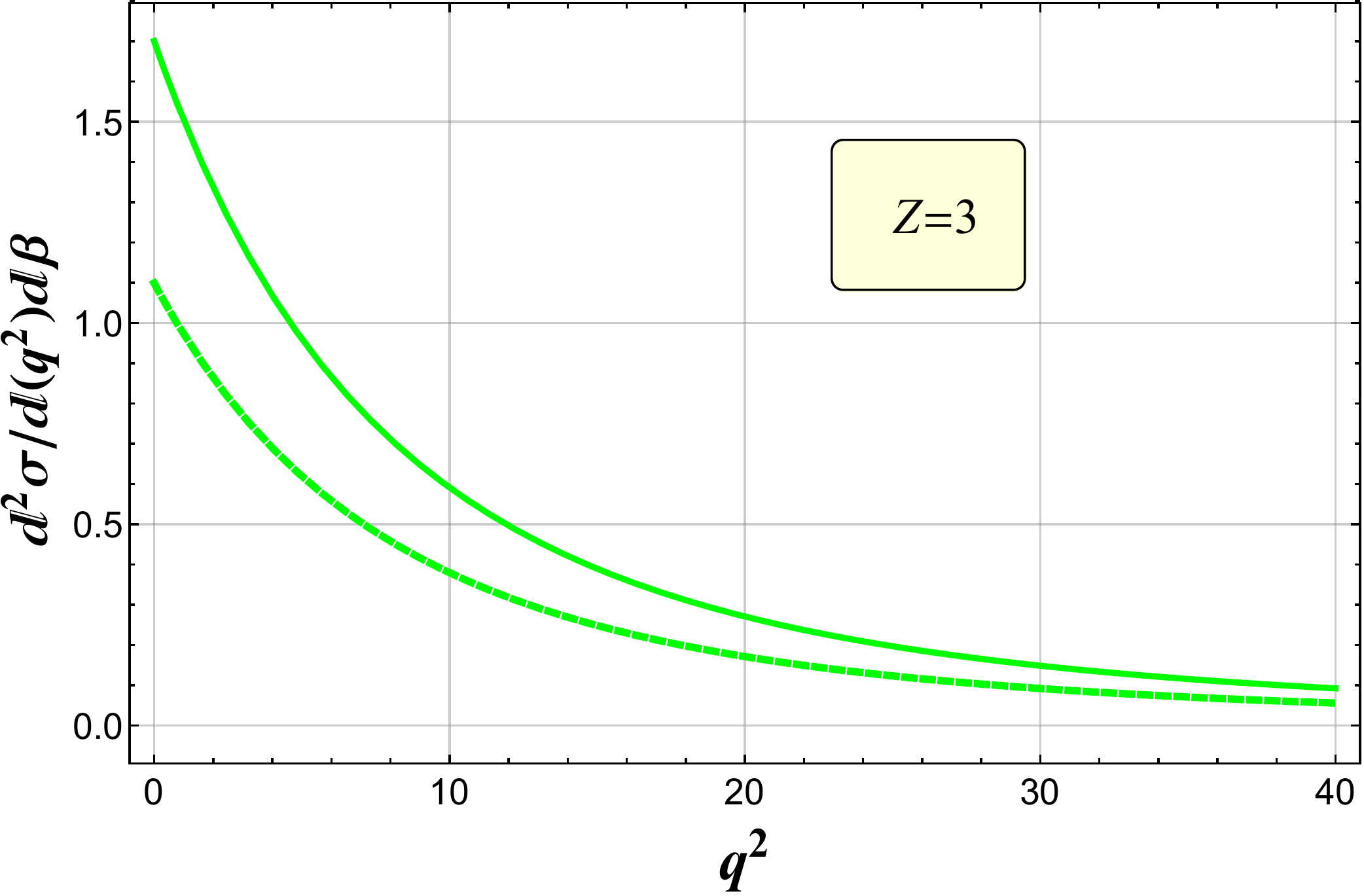}}\label{F1:c}
\subfigure{\includegraphics[width=0.495\textwidth]{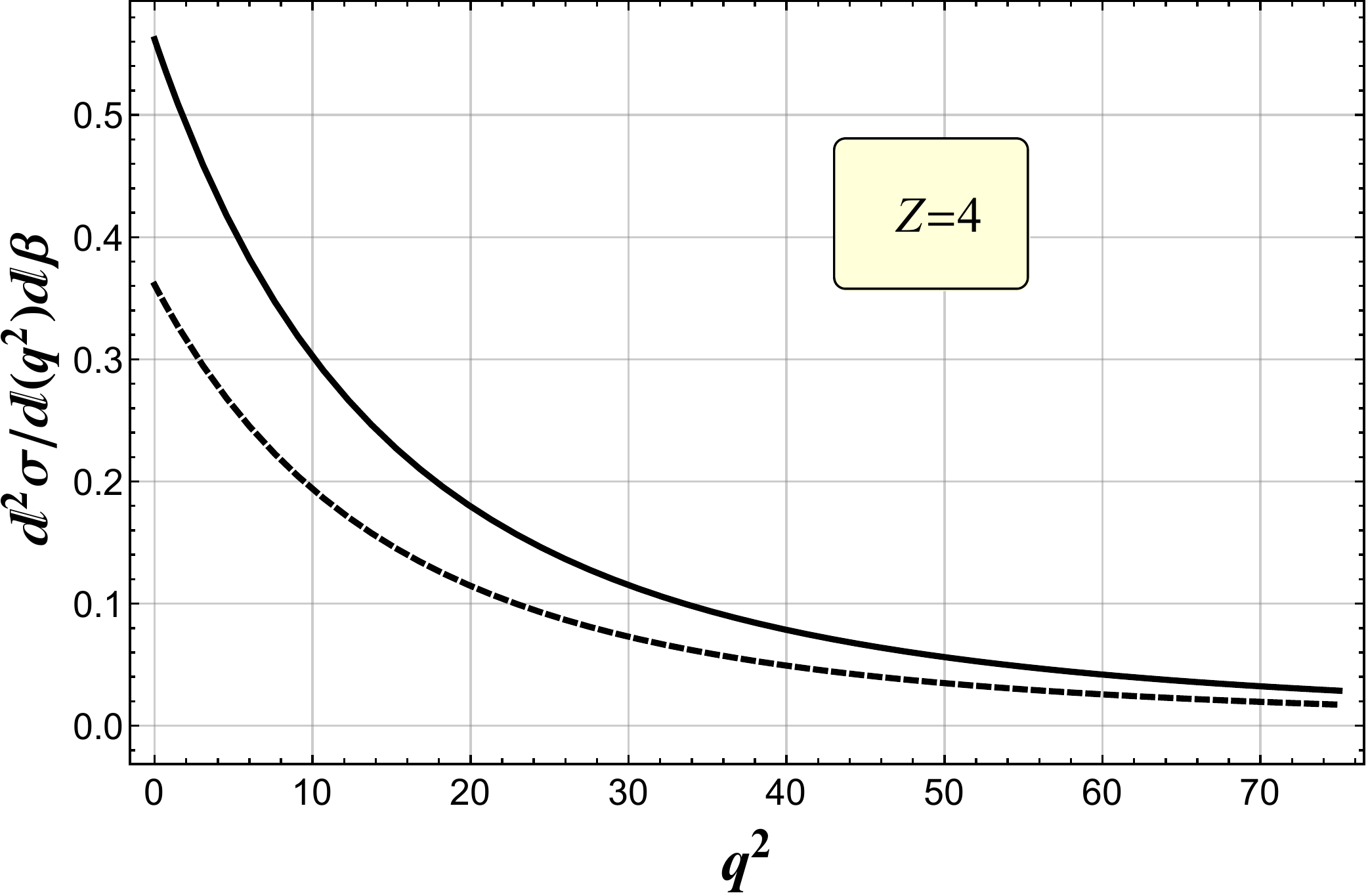}}\label{F1:d}
\subfigure{\includegraphics[width=0.495\textwidth]{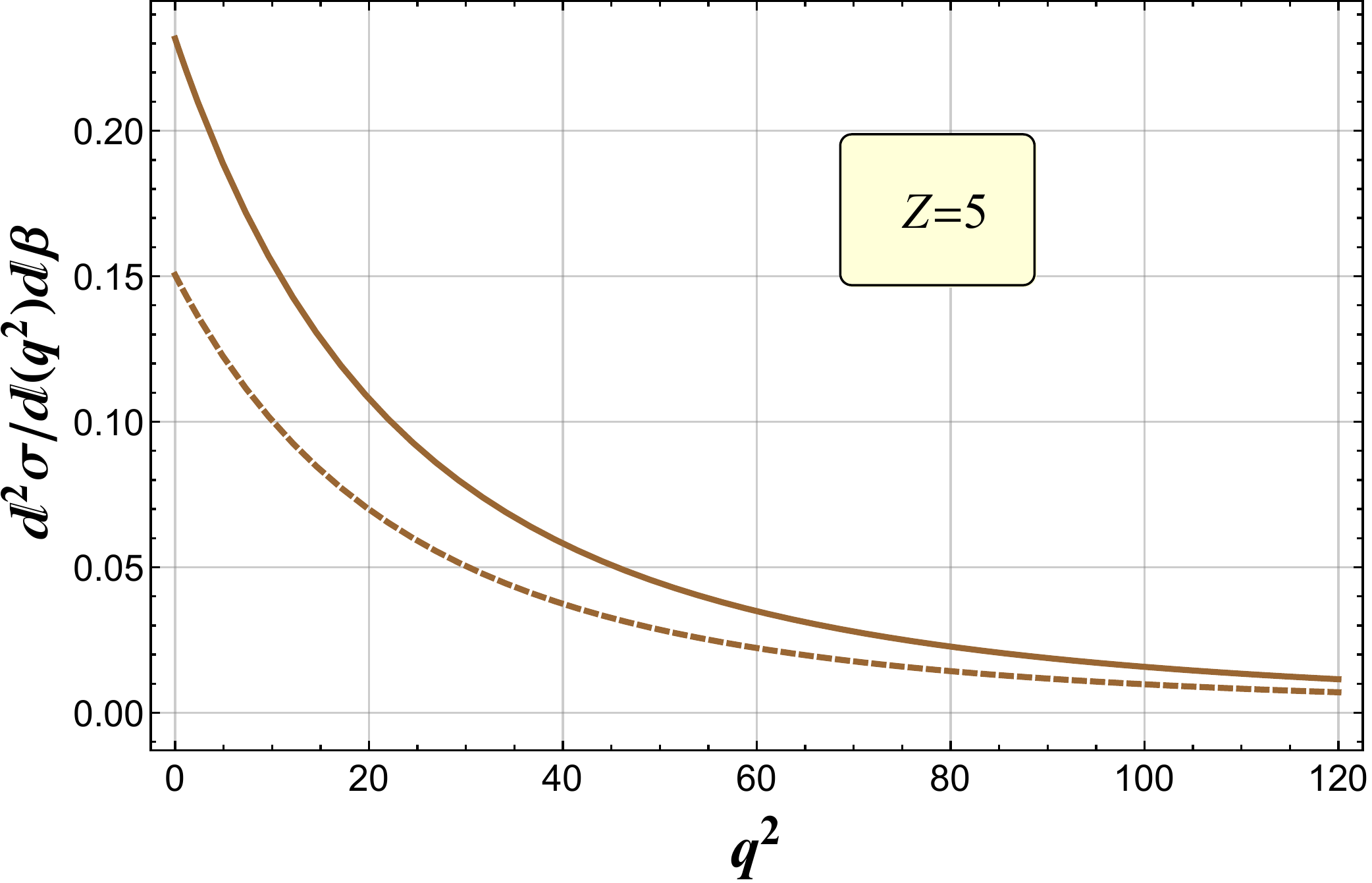}}\label{F1:e}
 \label{F1}
\end{figure}

\begin{figure}[htbp]
\centering
\caption{Distribution $d\sigma/d(q^2)$ in $10^{-10} a_0^4$ is presented as a function of $q^2$ in $a_0^{-2}$, where
$a_0$ is the Bohr radius. The solid lines correspond to the photon energies $\omega=145,800,2000,3750,6100$ eV, whereas the dashed lines correspond to the photon energies $\omega=180,1000,2500,4700,7600$ eV for the helium-like atoms with $Z=1;2;3;4;5$, respectively.}
\subfigure{\includegraphics[width=0.495\textwidth]{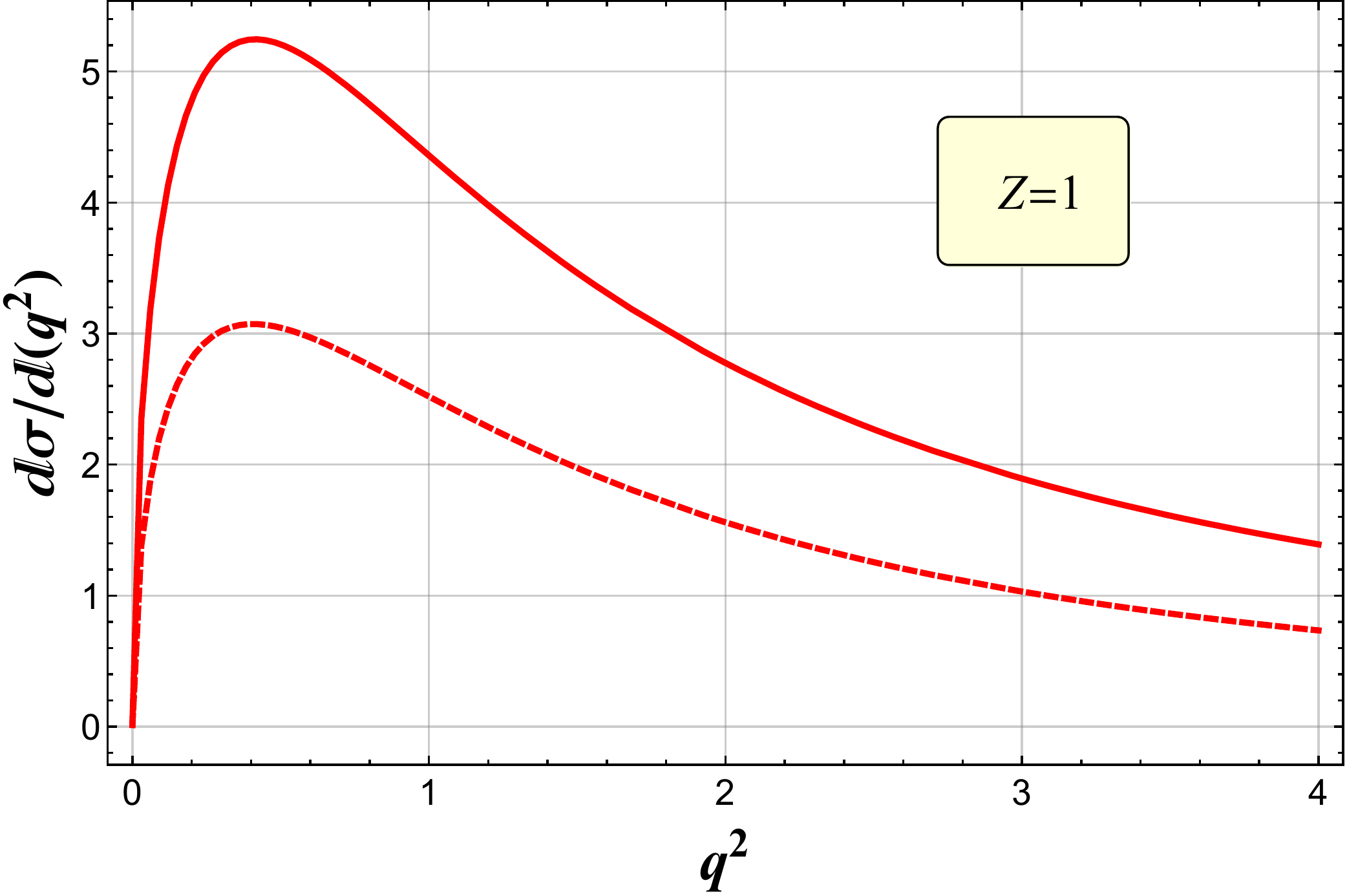}}\label{F4:a}
\subfigure{\includegraphics[width=0.495\textwidth]{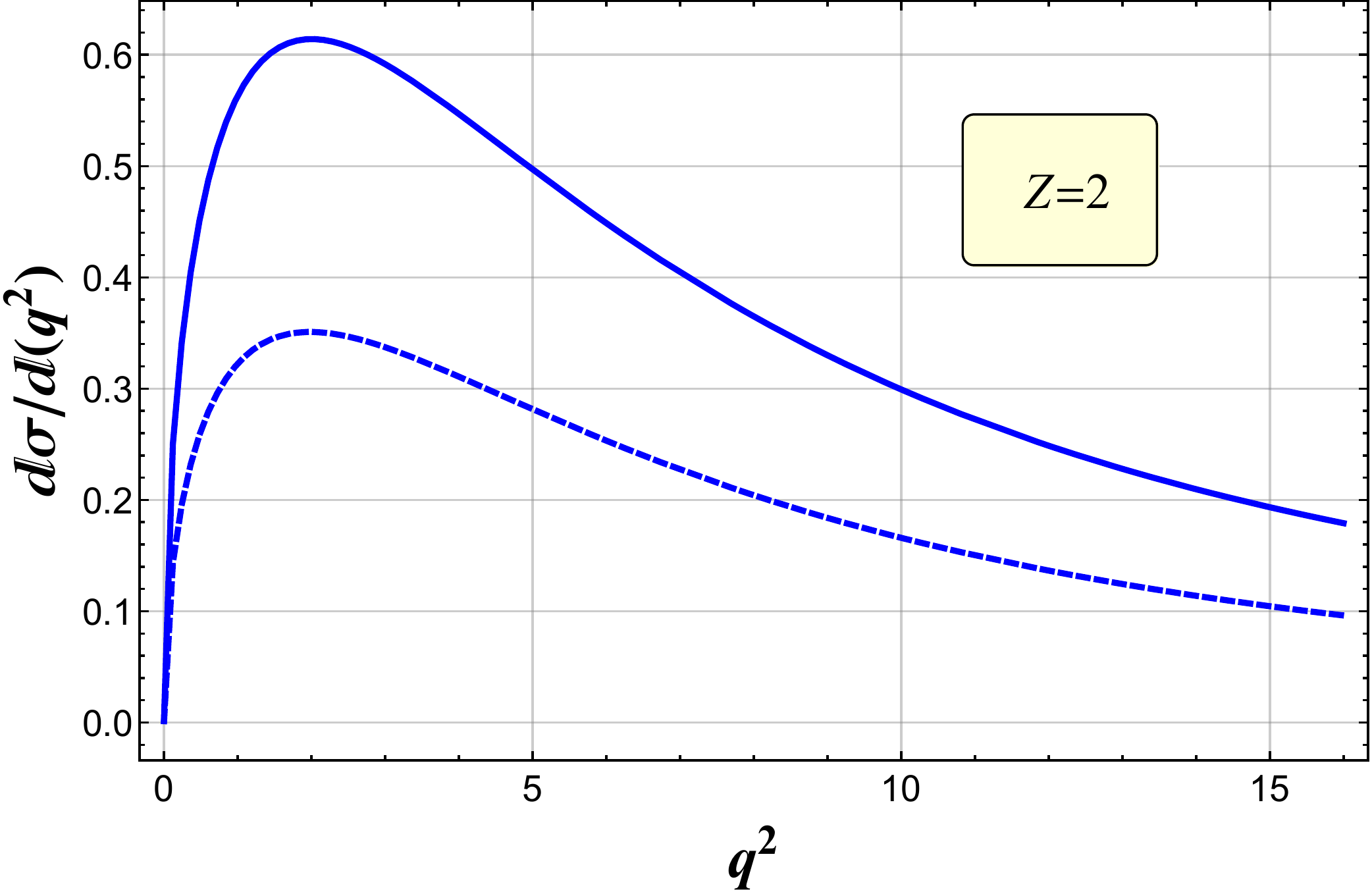}}\label{F4:b}
\subfigure{\includegraphics[width=0.495\textwidth]{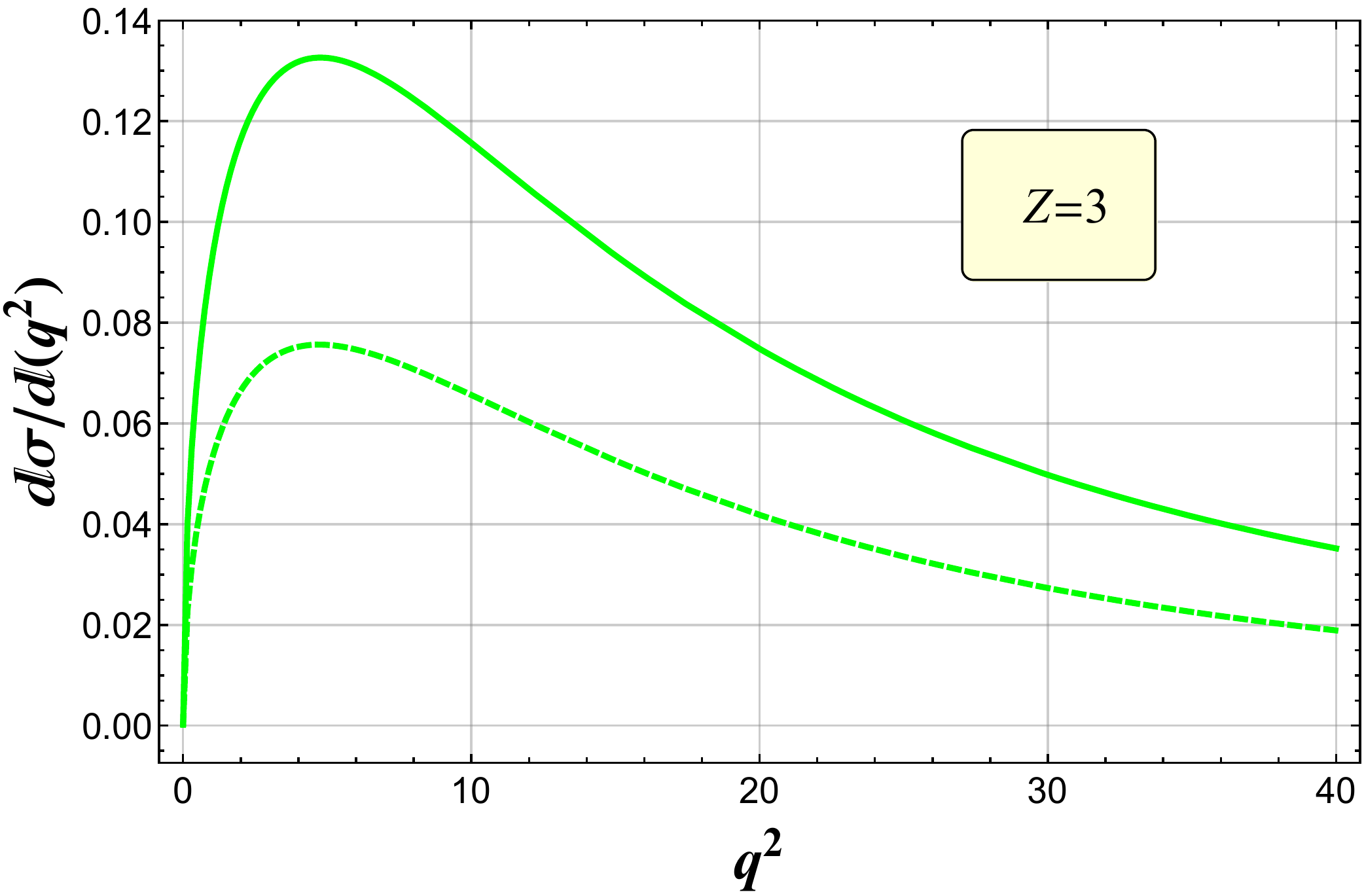}}\label{F4:c}
\subfigure{\includegraphics[width=0.495\textwidth]{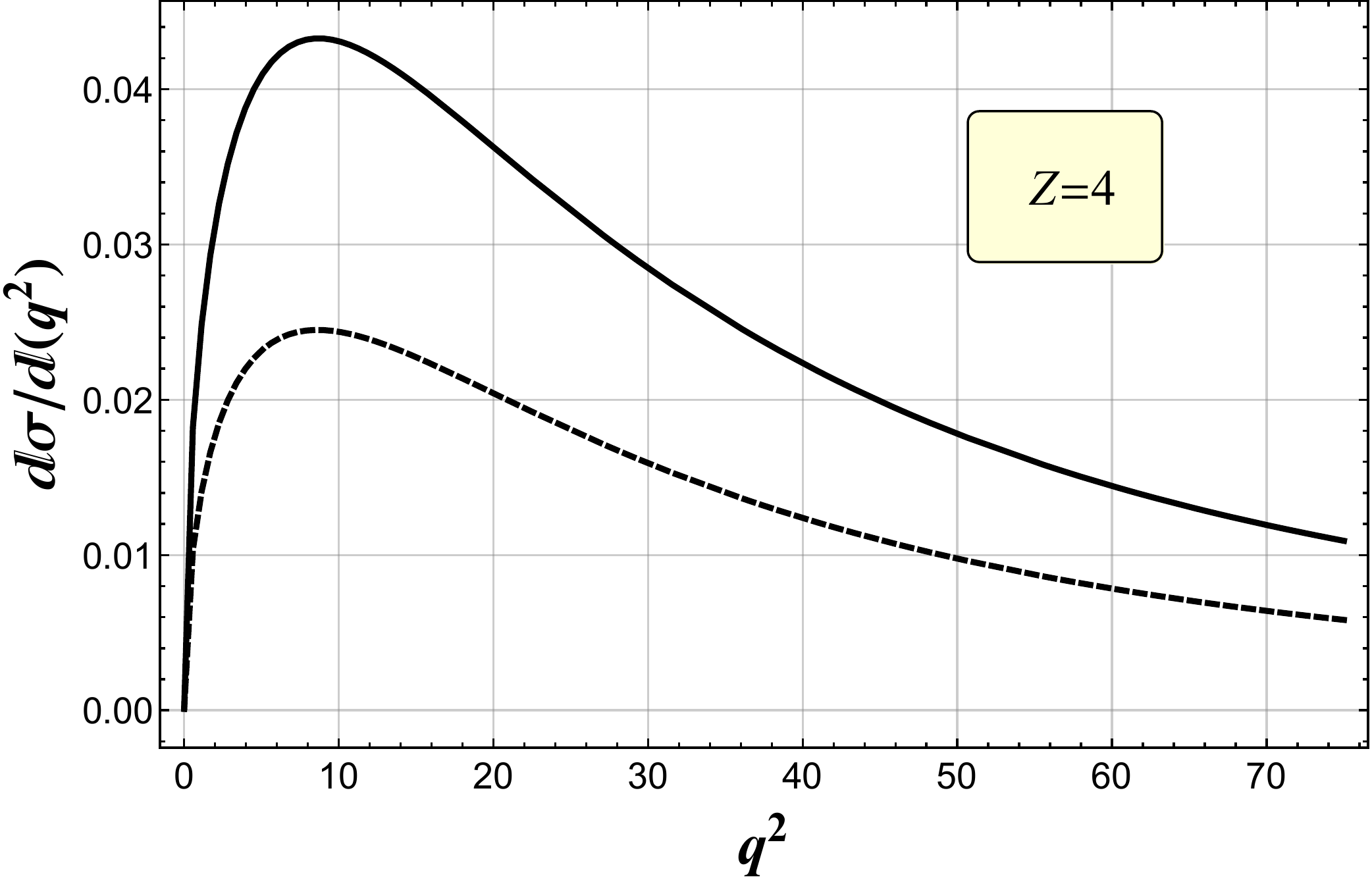}}\label{F4:d}
\subfigure{\includegraphics[width=0.495\textwidth]{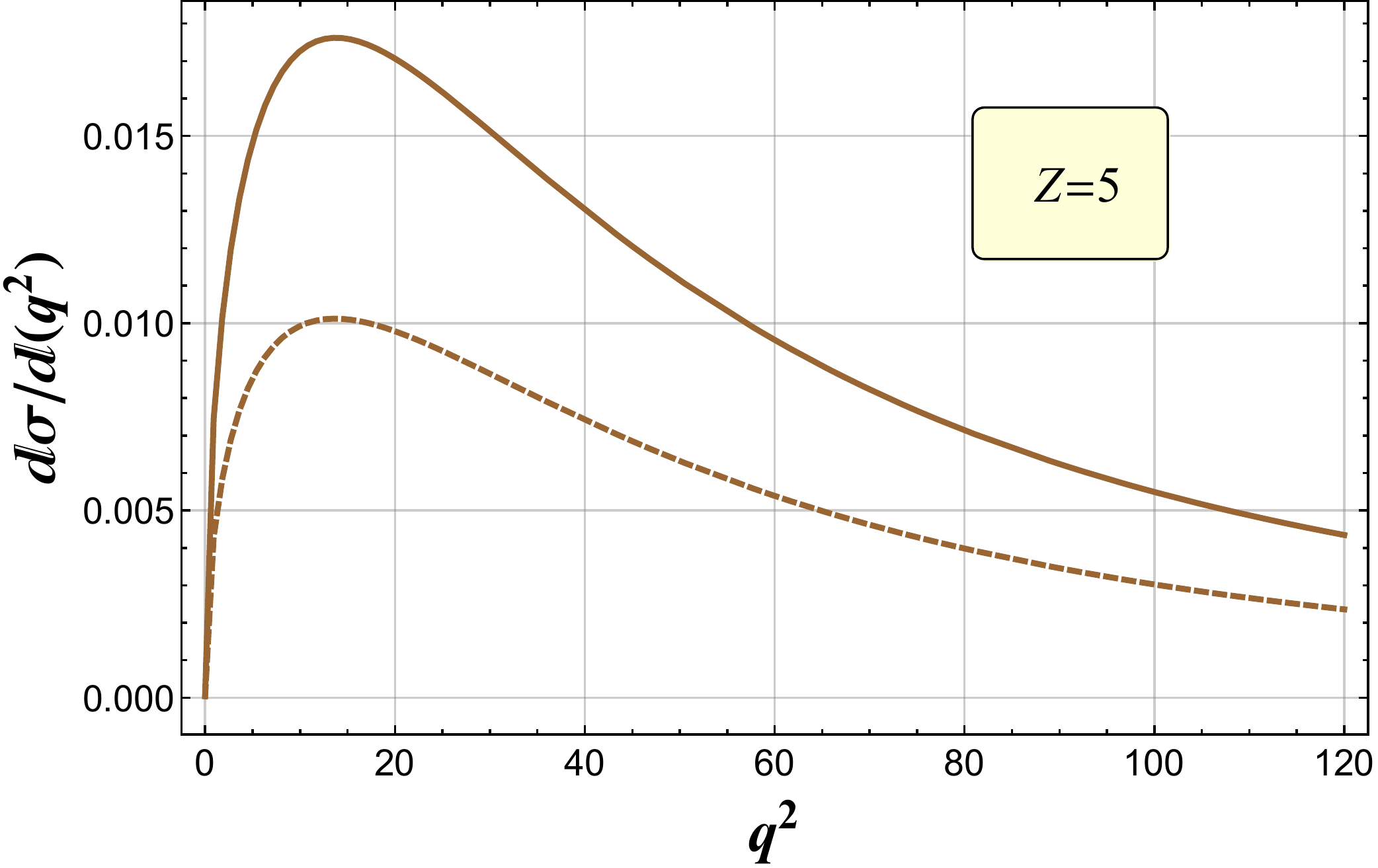}}\label{F4:e}
 \label{F2}
\end{figure}

As is seen from Fig.\ref{F1}, at equal photoelectron energies the magnitude of the double differential cross section  rapidly decreases with recoil momentum growth. The magnitude of it is the smaller the bigger is the nuclear charge. Fig.\ref{F2} presents the dependence of $d\sigma/d(q^2)$upon recoil moment. The curves for different $Z$ are similar having a profound maximum and rapidly decreasing in magnitude with increase of $Z$.

\begin{figure}[htbp]
\caption{Distribution $d^2\sigma/d\tau d\beta$ in barns is presented as a function of $\tau=(\textbf{p}_1\cdot \textbf{p}_2)/(p_1 p_2)$ for $\beta=0$. The curves on the plot $(\textrm{a})$ correspond to the photon energies $\omega=145,800,2000,3750,6100$ eV, whereas the curves on the plot $(\textrm{b})$ correspond to the photon energies $\omega=180,1000,2500,4700,7600$ eV for $\textrm{H}^-$, $\textrm{He}$ and two-electron ions with
$Z=3;4;5$, respectively.}
\subfigure{\includegraphics[width=0.495\textwidth]{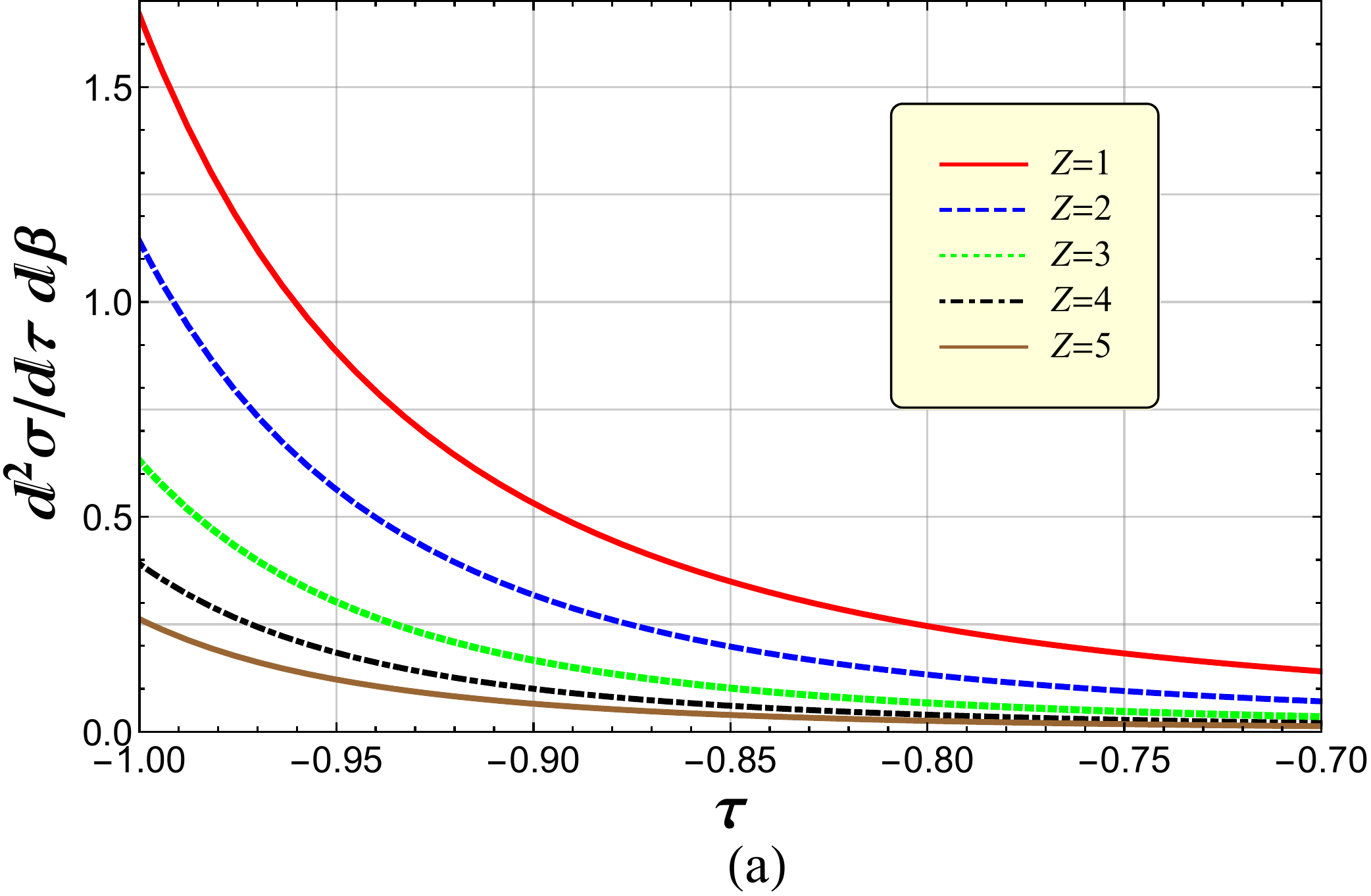}}\label{F2:a}
\subfigure{\includegraphics[width=0.495\textwidth]{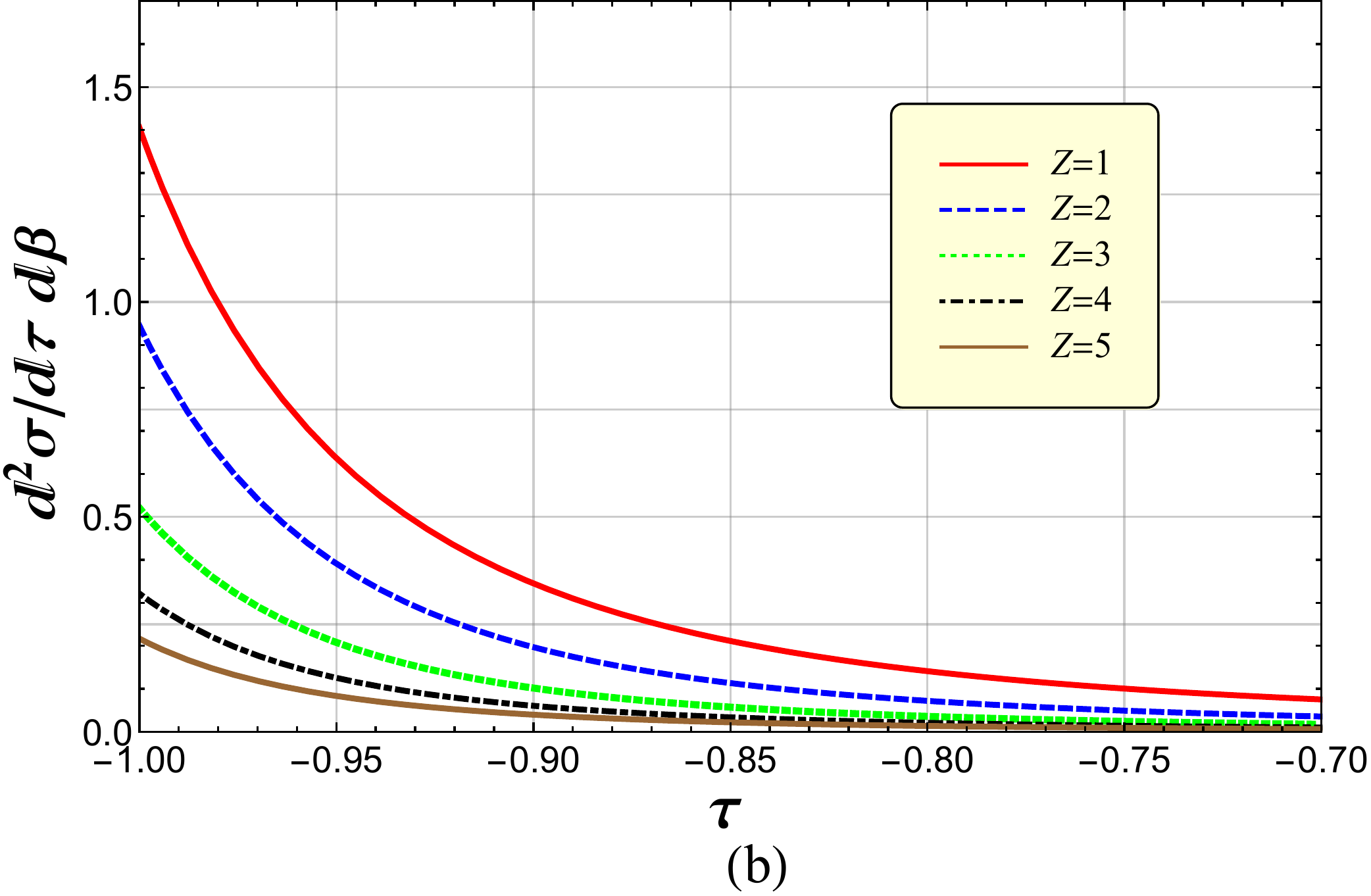}}\label{F2:b}
 \label{F3}
\end{figure}

\begin{figure}[htbp]
\caption{Distribution $d\sigma/d\tau$ in barns is presented as a function of $\tau$. The curves on the plot $(\textrm{a})$ correspond to the photon energies $\omega=145,800,2000,3750,6100$ eV, whereas the curves on the plot $(\textrm{b})$ correspond to the photon energies $\omega=180,1000,2500,4700,7600$ eV for $\textrm{H}^-$, $\textrm{He}$ and two-electron ions with
$Z=3;4;5$, respectively.}
\subfigure{\includegraphics[width=0.495\textwidth]{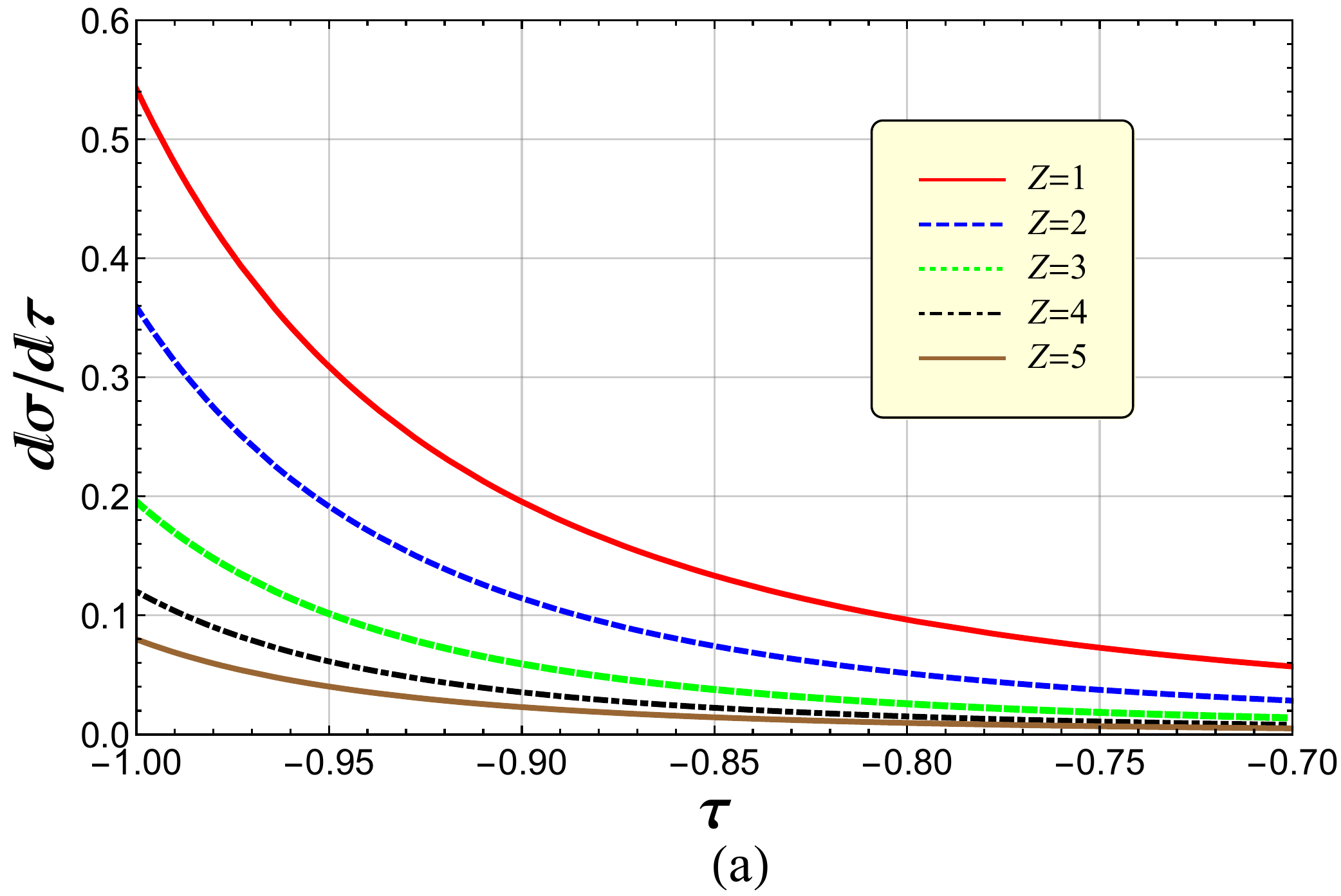}}\label{F3:a}
\subfigure{\includegraphics[width=0.495\textwidth]{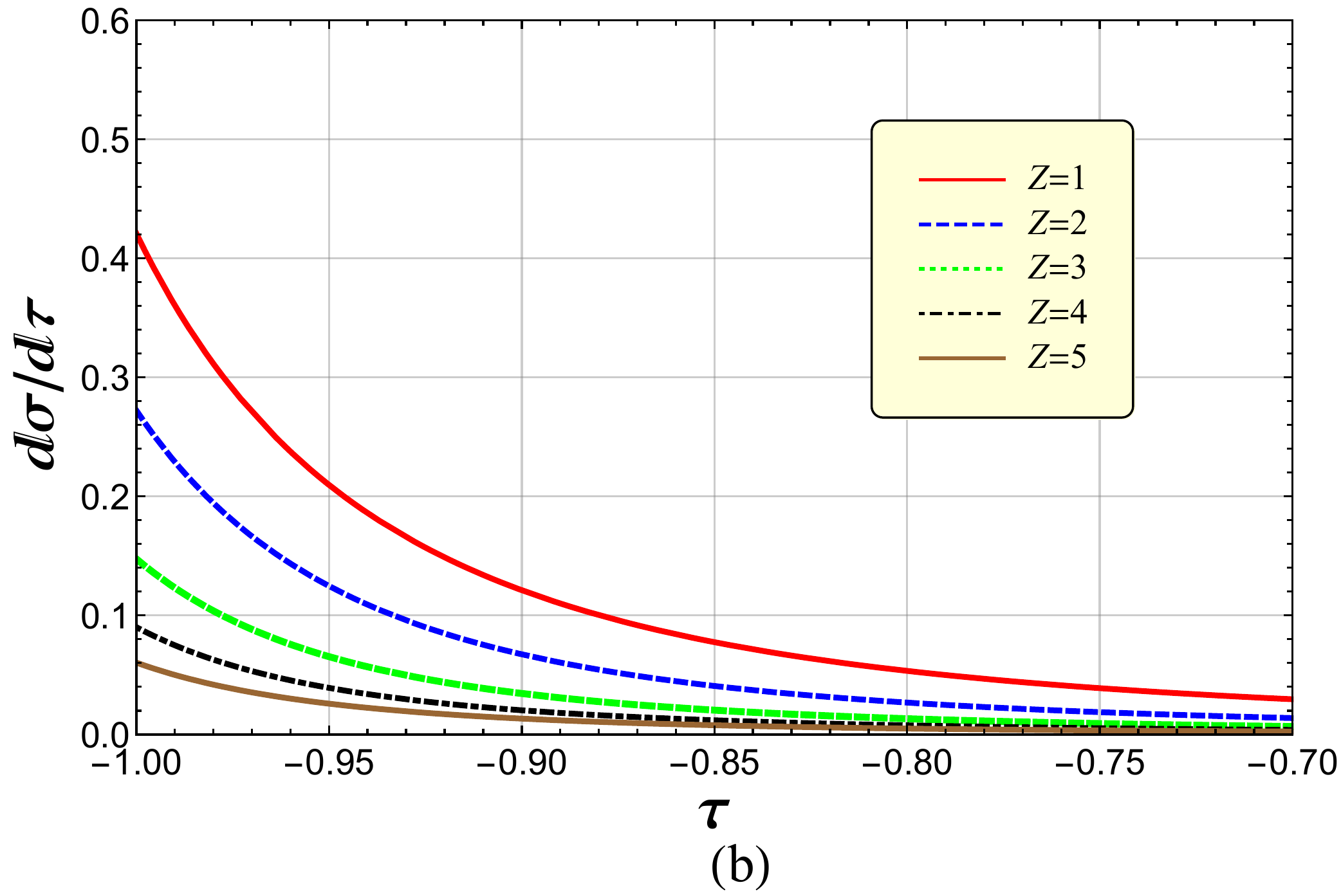}}\label{F3:b}
 \label{F4}
\end{figure}

In Figs.\ref{F3} and \ref{F4} we present the double differential distributions $d^2\sigma/d\tau d\beta$ at $\beta=0$ and the angular distributions $d\sigma/d\tau$ studied in \cite{8a} for $\textrm{He}$ atom, respectively.
This enables us to trace the $Z$ dependence of the effect. In Fig.\ref{F3} we compare also the results found by employing the functions on the coalescence line $\tilde{\Psi}(R,0)$ obtained in \cite{11}, \cite{12} with
those obtained by using approximate functions suggested in \cite{10}. One
can see that the difference is negligible for $\textrm{H}^{-}$ and for $\textrm{He}$ atom. It
increases with $Z$, remaining very small at least for $Z\leq 5$.
Fig \ref{F3} demonstrates that the QFM cross-sections are rapidly decreasing with increase of the angle between the outdoing electrons momenta, their magnitudes, as in Fig. \ref{F1} and Fig.\ref{F2}, rapidly decrease with $Z$ growth.
The results depicted in Fig.\ref{F4} demonstrate rapid decrease of $d\sigma/d\tau$ with decrease of the angle between the outgoing electrons momenta and growth of $Z$.
In Fig. 5(a-e) we present in details the photoelectron energy distributions $d\sigma/d\beta$ at $\beta=0$ for ($Z=1-5$).
Fig. 5 depicts the dependence of $d\sigma/d\beta$ upon photon energy, showing its rapid monotonic decrease with $\omega$. Note, however, that $d \sigma/d \beta$  is bigger for $Z=2$, than for $Z=1$, and the decrease on the way from $Z=2$ to $Z=5$ is relatively slow.

This can be useful for extension of the analysis carried out in \cite{8a} for another values of the photon energies and for other targets.

\section{Summary}

As it was mentioned above, the QFM predicted 45 years ago \cite{1} was
beyond the possibilities of experimental investigations for a long time. The
work \cite{8} provided experimental evidence of the existence of QFM. Recent
publication \cite{8a} provided experimental data on the double and single
differential distributions for $\textrm{He}$ atom and for $\textrm{H}_{2}$ molecule. This
enables us to hope that studies of QFM for other targets will take place
transforming a couple of experiments into whole domain of research that will present data on short range inter-electron correlations in a whole variety of systems of which $\textrm{He}$, $\textrm{H}^-$ and other helium-like ions form only a small domain.

The QFM is interesting from several points of view. It probes the wave
function between the bound electrons at small distances and provides a good
test for the wave functions at the electron-\texttt{electron} coalescence line. The QFM
depends on the proper inclusion of correlations of the bound state
electrons. It can not be reproduced by uncorrelated bound state functions
\cite{7}. The QFM is the only mechanism of ionization which requires going
beyond the dipole approximation since it takes place only if the quadrupole
terms in photon-electron interaction are included.

This stimulated us to calculate various characteristics of the double
photoionization for the negative ion $\textrm{H}^{-}$, $\textrm{He}$\ atom and for
two-electron ions $\textrm{Li}^{+},\textrm{Be}^{++}$ and $\textrm{B}^{+3}$ with $Z=3,4,5,$ respectively
in the region of QFM domination at the photon energies $I\ll \omega \ll \mu $.
We trace the $Z$ dependence for the double differential distribution.
For $\textrm{He}$ we traced the dependence of the photoelectron energy
distribution on the photon energy. Since the interest to the QFM renewed
recently\cite{8a}, \cite{13} we hope these data to be useful.

\appendix

\section{}\label{SA}

In this Appendix we present the refined formula for calculation of the three-dimensional integral $S_1(q)$ defined by Eq.(\ref{23}).

Inserting representations (\ref{21}) into the RHS of Eq.(\ref{23}), we obtain
\begin{equation} \label{A1}
S_{1}(q)\ = N(p_1)N(p_2)\int d^{3}R~e^{i\mathbf{q}\mathbf{R}}~
_{1}F_{1}(i\xi _{1},1,ip_{1}r-i\mathbf{p}_1\mathbf{r})_{1}F_{1}(i\xi _{2},1,ip_{2}r-i\mathbf{p}_2\mathbf{r})
\tilde{\Psi}(R,0),
\end{equation}
where $\tilde{\Psi}(R,0)$ represents the two-electron wave function (in the ground state) at the electron-electron coalescence line. The integral (\ref{A1}) can be easily calculated for $\tilde{\Psi}(R,0)$ presented in the form
\begin{equation} \label{A2}
\tilde{\Psi}(R,0)=\sum_{j=1}^n C_j \exp(-\lambda_j R).
\end{equation}
The Pekeris-like wave functions which we applied \cite{11,12} do not have the form (\ref{A2}) at the electron-electron coalescence line. However, fortunately, it is sufficient to include five separate exponential terms ($n=5$) to obtain extremely  accurate  wave function $\tilde{\Psi}(R,0)$ of the form (\ref{A2}) by fitting the Pekeris-like wave functions with the number of shells $\Omega=25$ \cite{12}.

It follows from Eqs.(\ref{A1}) and (\ref{A2}) that calculations of the integral (\ref{A1}) reduce to computation of the integral
\begin{equation} \label{A3}
I(q,\lambda,s) =\int e^{i\mathbf{q}\mathbf{R}-\lambda R}~
_{1}F_{1}(i\xi _{1},1,ip_{1}r-i\mathbf{p}_1\mathbf{r})_{1}F_{1}(i\xi _{2},1,ip_{2}r-i\mathbf{p}_2\mathbf{r}) R^s d^{3}R .
\end{equation}
It is clear that the analytic form for the latter integral with $s=0$ can be obtain by differentiation of the integral (\ref{A3}) with $s=-1$, in respect to parameter $\lambda$.
The analytic form of the integral $I(q,\lambda,-1)$ was derived in Ref.\cite{14}.
Now we employing this result  and take into account that integral (\ref{A1}) depends, in fact, only on $q^2$.
The evaluation mentioned above provides the required integral in the form:
\begin{eqnarray} \label{A4}
I(q,\lambda,0) =
~~~\nonumber~~~~~~~~~~~~~~~~~~~~~~~~~~~~~~~~~~~~~~~~~~~~~~~~~~~~~~~\\
-4\pi\left(\lambda^2+q^2\right)^{i(\xi_1+\xi_2)-1}
(p_2-p_1-i \lambda)^{-i \xi_1}(p_1-p_2-i \lambda)^{-i \xi_2}(p_1+p_2+i \lambda)^{-i (\xi_1+\xi_2)}\times
~~\nonumber~~~~~\\
\left\{_1F_1\left[i \xi_1+1,i \xi_2+1;2;h(q,\lambda)\right]
\frac{2\lambda \xi_1 \xi_2 h(q,\lambda)}{(p_1-p_2)^2+\lambda^2}~+~_1F_1\left[i \xi_1,i \xi_2;1;h(q,\lambda)\right]\times
\right.
~~~\nonumber~~~~~~\\
\left.
\left(\frac{\xi_1+\xi_2}{p_1+p_2+i\lambda}+\frac{\xi_1}{p_1-p_2+i\lambda}+\frac{\xi_2}{p_2-p_1+i\lambda}+
\frac{2\lambda[i(\xi_1+\xi_2)-1]}{\lambda^2+q^2}\right)
\right\},~\nonumber~~~
\end{eqnarray}
where
\begin{equation} \label{A5}
h(q,\lambda)=1-\frac{\lambda^2+q^2}{(p_1-p_2)^2+\lambda^2}.
\end{equation}

\newpage



\begin{figure}[htbp]
\centering
\caption{Distribution $d\sigma/d\beta$ in $10^{-10} a_0^2$ is shown as a function of the photon energy $\omega$  (in eV) for the helium-like isoelectronic sequence with $1\leq Z\leq5$ ($a_0$ is the Bohr radius).
Only the short ranges of $\omega$ containing the photon energies considered in the paper are presented.}
\subfigure{\includegraphics[width=0.495\textwidth]{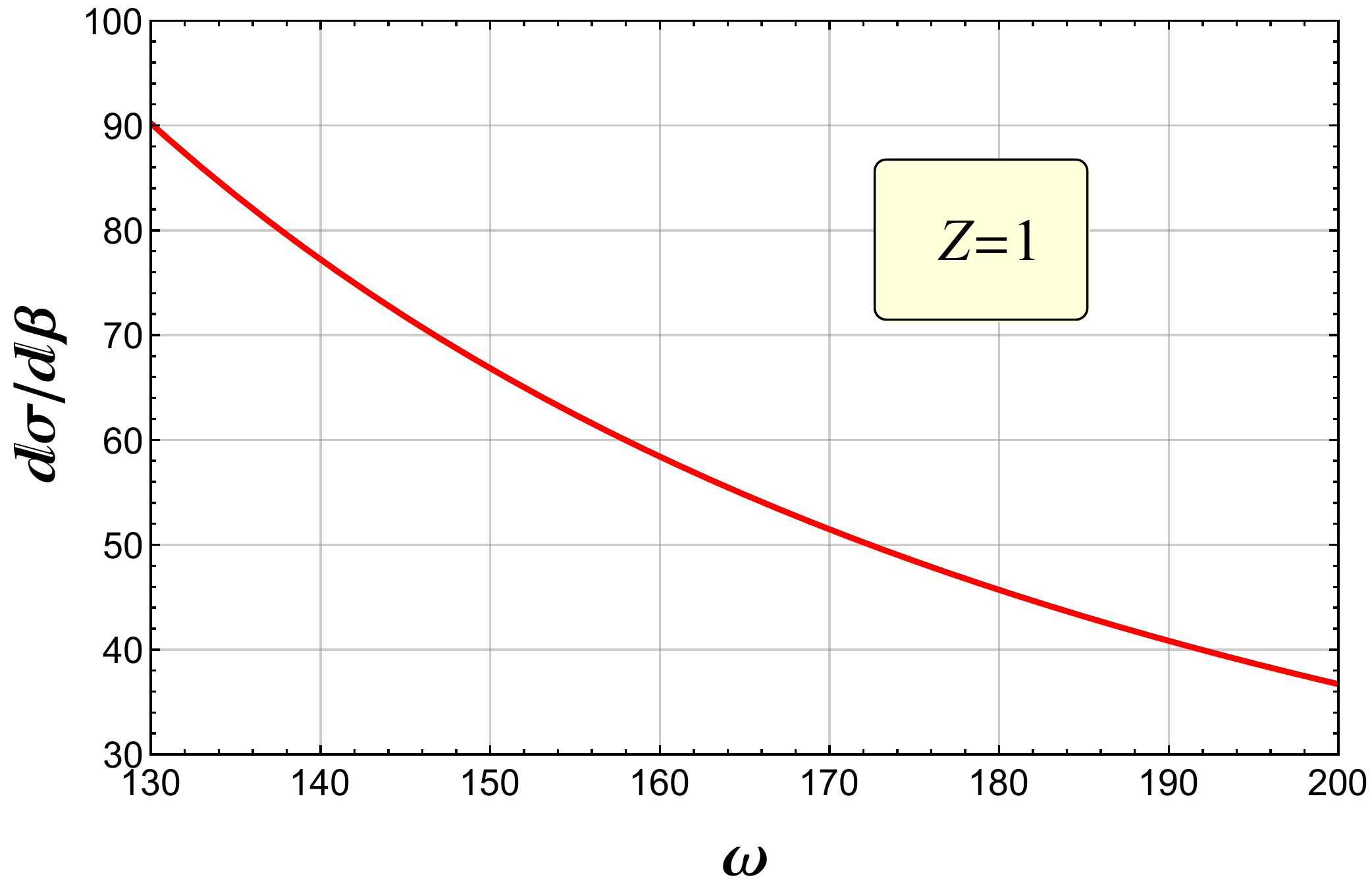}}\label{F6:a}
\subfigure{\includegraphics[width=0.495\textwidth]{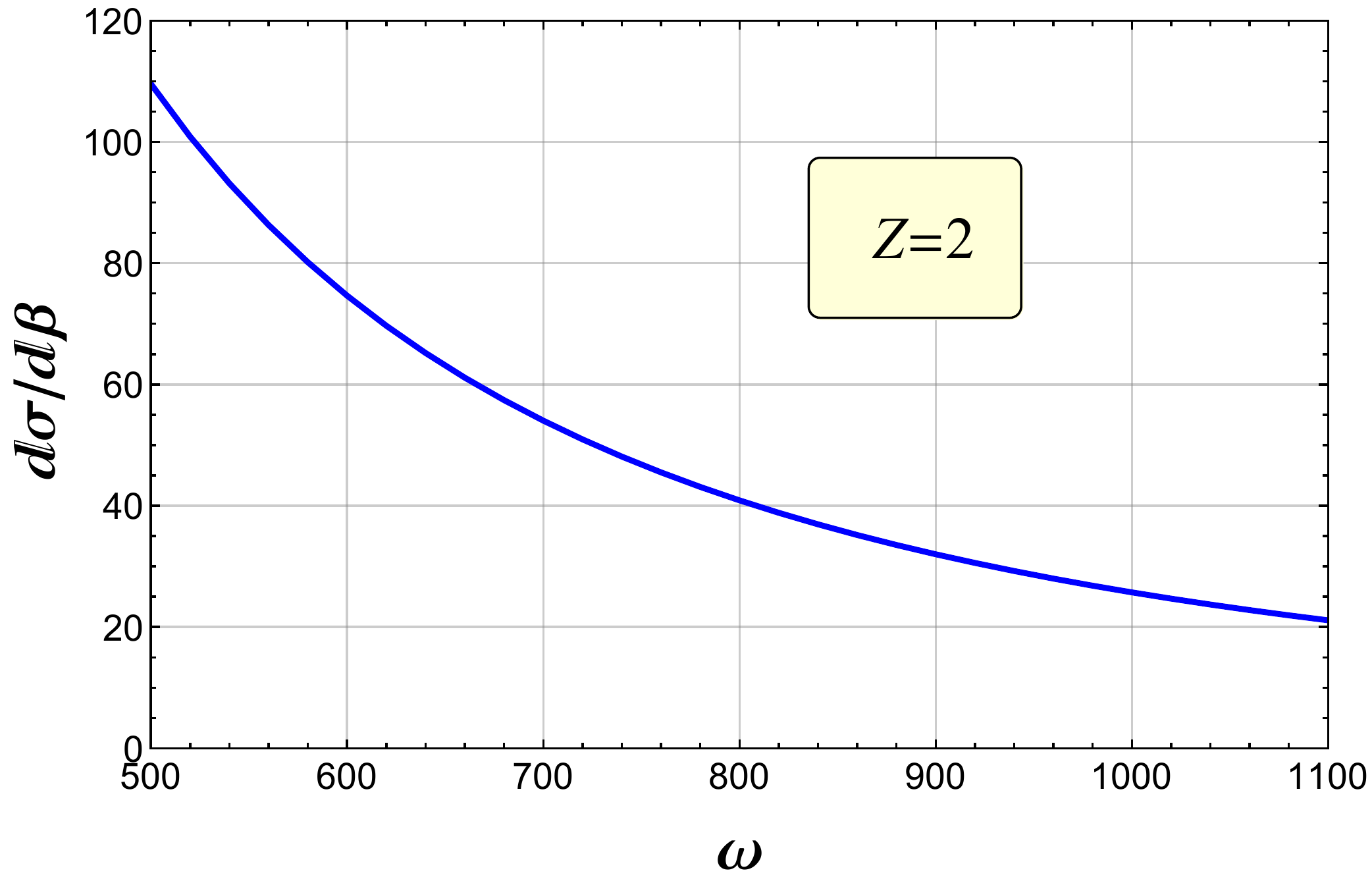}}\label{F6:b}
\subfigure{\includegraphics[width=0.495\textwidth]{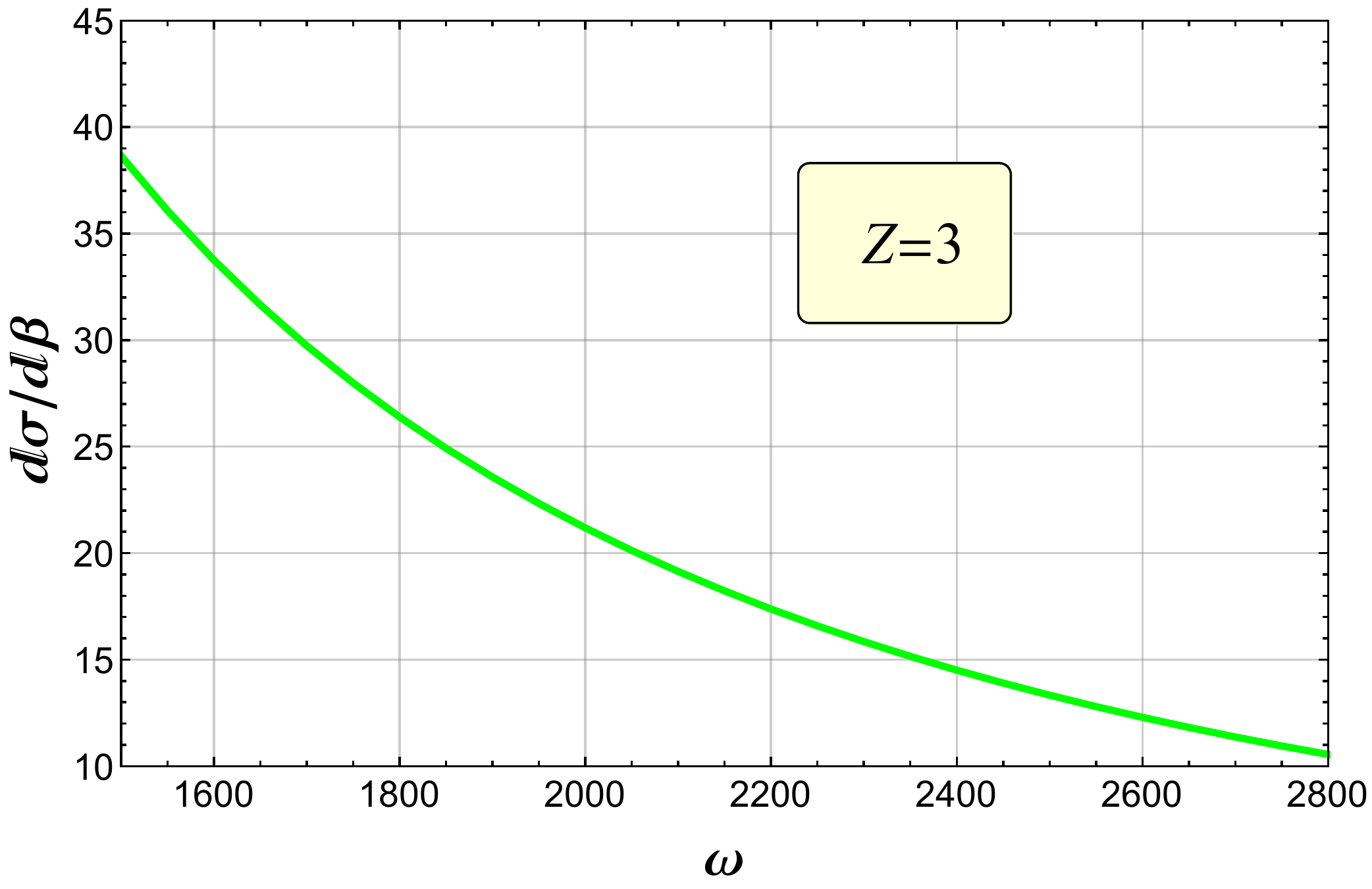}}\label{F6:c}
\subfigure{\includegraphics[width=0.495\textwidth]{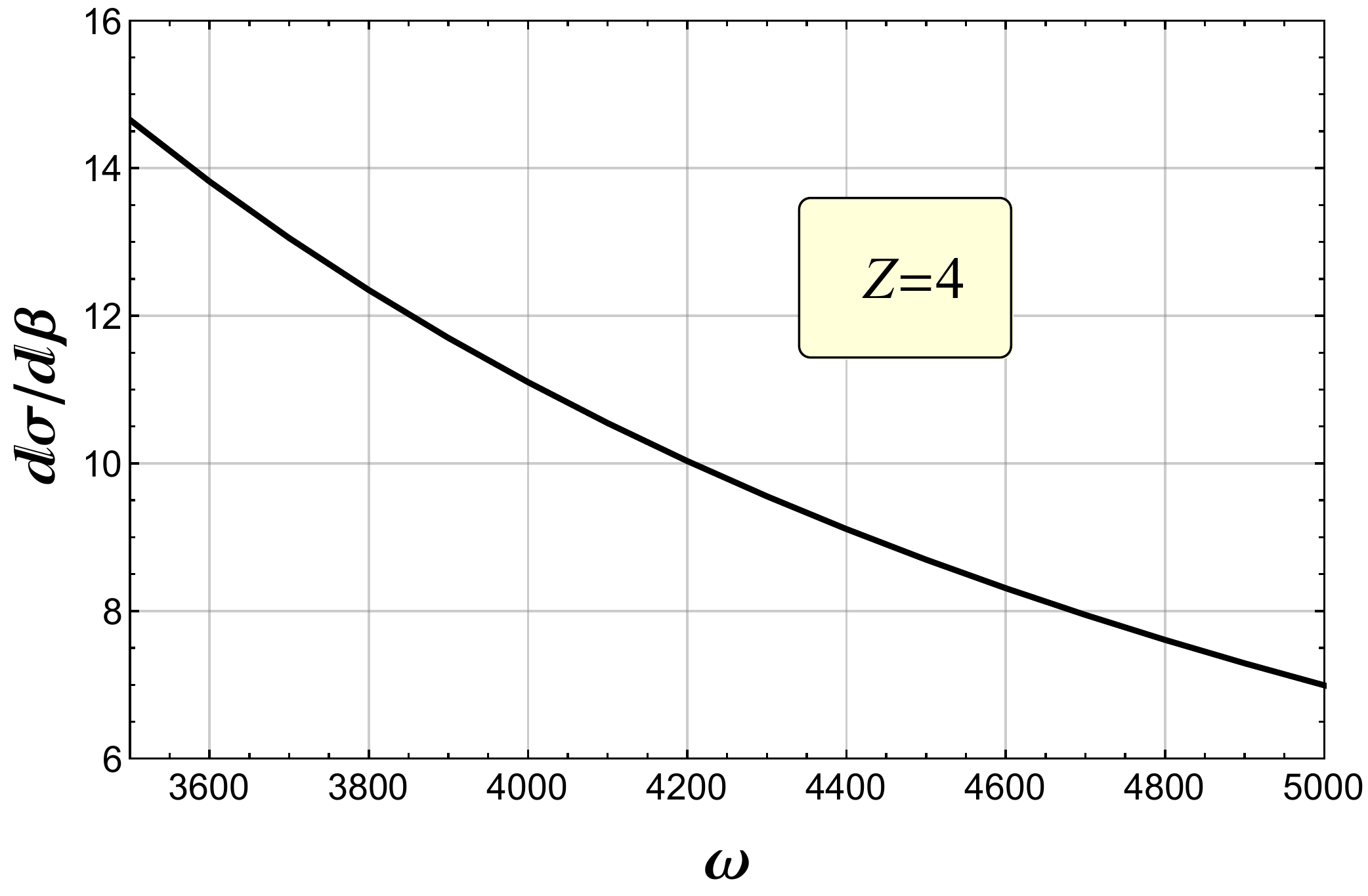}}\label{F6:d}
\subfigure{\includegraphics[width=0.495\textwidth]{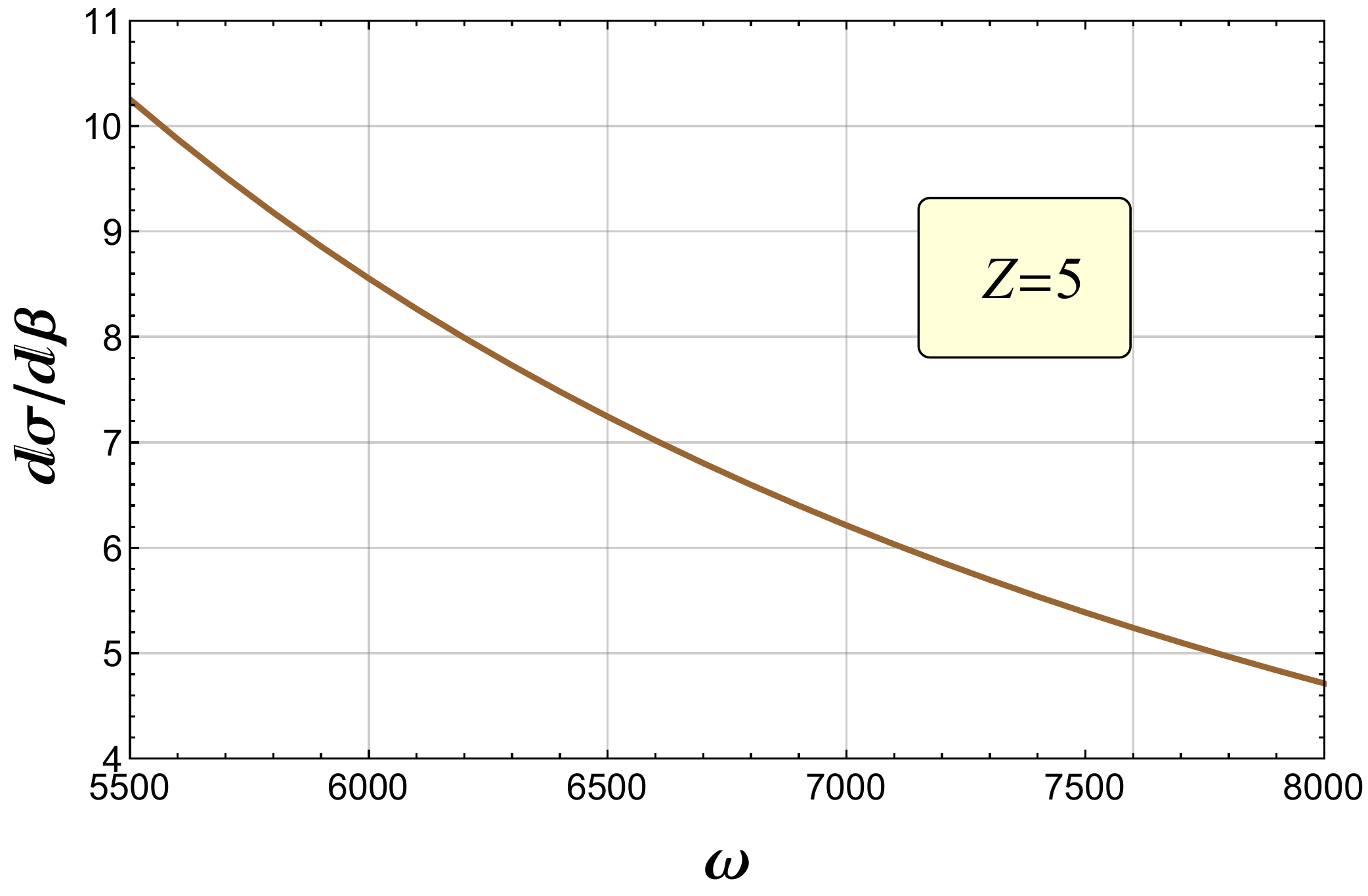}}\label{F6:e}
 \label{F5}
\end{figure}


\begin{thebibliography}{99}
\bibitem{1} M. Ya. Amusia, E. G. Drukarev, V. G. Gorshkov, M,P.Kazachkov,J.
Phys. B \textbf{8},1248 (1975).

\bibitem{2} E. G. Drukarev and F. F. Karpeshin, J. Phys. B. \textbf{9}, 399
(1976).

\bibitem{3} R. Krivec, M. Ya. Amusia, and V. B. Mandelzweig, Phys. Rev. A
\textbf{64}, 052708 (2001).

\bibitem{4} T. Suri\'{c}, E. G. Drukarev and R. H. Pratt, Phys. Rev. A%
\textbf{67}, 022709 (2003).

\bibitem{5} T. Kato, Commun. Pure Appl. Math \textbf{10}, 151 (1957).

\bibitem{6} E. Z. Liverts, M. Ya. Amusia, E. G. Drukarev, R. Krivec and V.
B. Mandelzweig, Phys. Rev. A \textbf{71}, 012715 (2005).

\bibitem{7} E. G. Drukarev and A. I. Mikhailov \emph{High Energy Atomic
Physics}, Springer International Publishing Switzerland 2016.

\bibitem{8} M. S. Sch\"{o}ffler et al. Phys. Rev. Lett. \textbf{111},
0132003 (2013).

\bibitem{8a} S. Grundmann, V. Serov, F. Trinter, K. Fehre, N. Strenger, A.
Pier, M. Kircher, D. Trabert, M. Weller, L. L. Kaiser, A. W. Bray, L. Ph. H.
Schmidt, J. B. Williams, T. Jahnke, R. Dorner, M. S. Schoffler, and A. S.
Kheifets Physical Review Letters, submitted (2020); arXiv:
https://arxiv.org/abs/2001.07713

\bibitem{9} M. Ya. Amusia, E. G. Drukarev, E. Z. Liverts, and A. I.
Mikhailov, Phys. Rev. A \textbf{87}, 043423 (2013).

\bibitem{10} E. Z. Liverts, M. Ya. Amusia, R. Krivec and V. B. Mandelzweig,
Phys. Rev. A \textbf{73}, 012514 (2006).

\bibitem{11} E. Z. Liverts and N. Barnea, Compt. Phys. Comm. \textbf{182},
1790 (2012).

\bibitem{12} E. Z. Liverts and N. Barnea, Compt. Phys. Comm. \textbf{184},
2596 (2013).

\bibitem{13} S.-G. Chen, W.-C. Jiang, S. Grundmann, F. Trinter, M. S.
Schoffler, T. Jahnke, R. Dorner, H. Liang, M.-X. Wang, L.-Y. Peng, and Q.
Gong, Phys. Rev. Lett. \textbf{124}, 043201 (2020).

\bibitem{14} A. Nordsieck, Phys. Rev. \textbf{93}, 785 (1954).

\end{thebibliography}
\end{document}